\documentclass[sigconf]{acmart}

\usepackage{listings}
\usepackage{multirow}

\usepackage{amsmath}
\usepackage{bm} 
\usepackage{amsfonts}

\AtBeginDocument{%
  \providecommand\BibTeX{{%
    \normalfont B\kern-0.5em{\scshape i\kern-0.25em b}\kern-0.8em\TeX}}}

\setcopyright{none}
\renewcommand\footnotetextcopyrightpermission[1]{} 

\begin{document}

\title{Optimizing Tensor Programs on Flexible Storage}

\author{Maximilian Schleich}
\affiliation{%
  \institution{RelationalAI}
  \country{USA}
}

\author{Amir Shaikhha}
\affiliation{%
  \institution{University of Edinburgh}
  \country{United Kingdom}}

\author{Dan Suciu}
\affiliation{%
  \institution{University of Washington}
  \country{USA}
}

\definecolor{forestgreen}{rgb}{0.13, 0.55, 0.13}
\colorlet{myblue}{blue!70!black}
\colorlet{mygreen}{green!70!black}
\colorlet{mypurple}{purple!70!black}
\colorlet{myorange}{orange!70!black}
\colorlet{myred}{red!70!black}
\colorlet{myyellow}{yellow!60!black}

\lstdefinelanguage{sdql}{
  morekeywords={if,then,else,let,in,not,%
  true,false,subarr,%
  sum,range,merge,%
  CREATE,TENSOR,AS,WITH,SCALAR,ARRAY,SIZE,%
  HASHMAP,TRIE,%
  int,real,dense_int,bool,string},%
  sensitive,%
  morecomment=[l]//,%
  morecomment=[s]{/*}{*/},%
  morestring=[b]",%
  morestring=[b]',%
  showstringspaces=false,%
  breaklines=true,%
  mathescape=true,%
  showspaces=false,
  showtabs=false,
  showstringspaces=false,
  breakatwhitespace=true,
  xleftmargin=1em,
  aboveskip=1pt,
  belowskip=1pt,
  lineskip=-0.2pt,
  basicstyle=\small\ttfamily\color{white!15!black},
  keywordstyle=\small\ttfamily\bfseries\color{myblue},%
  columns=fullflexible,
 commentstyle=\color{forestgreen},
  escapeinside={(*@}{@*)}
}[keywords,comments,strings]%

\lstset{language=sdql}
\lstMakeShortInline[columns=fixed, keepspaces=true, language=sdql]!

\newcommand{\transto}{\text{ $\leadsto$ }}
\newcommand{\transbi}{\text{ $\leftrightarrow$ }}

\newcommand{\system}{STOREL}
\newcommand{\lang}{SDQLite}
\newcommand{\taco}{Taco}
\newcommand{\scipy}{SciPy}
\newcommand{\numpy}{NumPy}
\newcommand{\tensorflow}{TensorFlow}
\newcommand{\pytorch}{PyTorch}
\newcommand{\duckdb}{DuckDB}
\newcommand{\julia}{Julia 1.6.2}

\newcommand{\dan}[1]{\textcolor{red}{[[Dan: #1]]}}
\newcommand{\mjs}[1]{\textcolor{blue}{[Max: #1]}}
\newcommand{\amir}[1]{\textcolor{purple}{[Amir: #1]}}

\newcommand{\defeq}{\stackrel{\text{def}}{=}}

\newcommand{\floor}[1]{\left\lfloor #1 \right\rfloor}
\newcommand{\ceil}[1]{\left\lceil #1 \right\rceil}

\newcommand{\eat}[1]{}

\newcommand{\calH}{\mathcal H}
\newcommand{\calM}{\mathcal M}
\newcommand{\calN}{\mathcal N}
\newcommand{\calI}{\mathcal I}
\newcommand{\calJ}{\mathcal J}
\newcommand{\calV}{[n]}
\newcommand{\calE}{\mathcal E}
\newcommand{\calF}{\mathcal F}
\newcommand{\calD}{\mathcal D}
\newcommand{\calW}{\mathcal W}
\newcommand{\calT}{\mathcal T}
\newcommand{\calS}{\mathcal S}
\newcommand{\calZ}{\mathcal Z}
\newcommand{\N}{{\mathbb N}}
\newcommand{\R}{{\mathbb R}}
\newcommand{\Z}{{\mathbb Z}}

\newcommand{\set}[1]{\{#1\}}                    
\newcommand{\setof}[2]{\{{#1}\mid{#2}\}}        

\newcommand{\translatebegin}{$\llbracket$}
\newcommand{\translateend}{$\rrbracket$}
\newcommand{\translate}[1]{\translatebegin\text{#1}\translateend}

\newcommand{\code}[1]{\texttt{#1}}
\newcommand{\card}[1]{{#1}}
\newcommand{\cardscalar}{\card{s}}
\newcommand{\carddict}[2]{#2[#1]}
\newcommand{\cardstatic}[1]{\##1}
\newcommand{\cardpre}{\textit{card}}
\newcommand{\elempre}{\textit{$card_{elem}$}}
\newcommand{\sizepre}{\textit{$card_{size}$}}
\newcommand{\cardelempre}{\textit{elem}}
\newcommand{\cardsizepre}{\textit{size}}
\newcommand{\selpre}{\textit{sel}}
\newcommand{\costpre}{\textit{cost}}
\newcommand{\rulespace}{\hspace{0.05cm}}
\newcommand{\coef}[1]{\gamma_{#1}}
\newcommand{\rulevspace}{\vspace{0.1cm}}

\newcommand{\rulecondition}{\hspace{0.2cm}\hfill if }

\newcommand{\rangen}[1]{[#1)}

\newcommand{\smartpara}[1]{\noindent \textbf{#1.}}

\newcommand{\expmmm}{\textit{MMM}}
\newcommand{\expsmmm}{\textit{$\Sigma$MMM}}
\newcommand{\expbatax}{\textit{BATAX}}
\newcommand{\expttm}{\textit{TTM}}
\newcommand{\expmttkrp}{\textit{MTTKRP}}

\begin{abstract}
  Tensor programs often need to process large tensors (vectors,
  matrices, or higher order tensors) that require a specialized
  storage format for their memory layout.  Several such layouts have
  been proposed in the literature, such as the Coordinate Format, the
  Compressed Sparse Row format, and many others, that were especially
  designed to optimally store tensors with specific sparsity
  properties.  However, existing tensor processing systems require
  specialized extensions in order to take advantage of every new
  storage format.  In this paper we describe a system that allows
  users to define flexible storage formats in a declarative tensor
  query language, similar to the language used by the tensor program.
  The programmer only needs to write {\em storage mappings}, which
  describe, in a declarative way, how the tensors are laid out in main
  memory.  Then, we describe a cost-based optimizer that optimizes the
  tensor program for the specific memory layout.  We demonstrate
  empirically significant performance improvements compared to
  state-of-the-art tensor processing systems.
\end{abstract}

\maketitle

\pagestyle{plain}

%


\keywords{databases, dense data}

\section{Introduction}

\label{sec:intro}

Linear algebra and, more generally, tensor algebra is used in a wide
variety of domains, such as science, engineering, machine learning,
data analysis.  Tensors are natural generalizations of vectors and
matrices from 1 and 2 dimensions to arbitrary dimensions, and highly
optimized implementations of tensor algebra operations are available
today in several popular libraries, such as SciPy, PyTorch, Julia,
TensorFlow, or Matlab.  While these libraries are highly optimized for
individual operations, compound operations require users to create
temporary tensors, which often destroys the locality and may even lead
to out of memory errors, when the intermediate results are too large.
Such operations are frequently encountered in complex tensor programs,
or tensor kernels, terms that we will use interchangeably in this
paper.

Several domain specific languages have been proposed for expressing
and optimizing entire tensor programs.  Examples include
SystemML~\cite{DBLP:journals/pvldb/BoehmRHSEP18},
TVM~\cite{DBLP:conf/osdi/ChenMJZYSCWHCGK18}, Halide~\cite{halide},
Taco~\cite{taco}, TASO~\cite{DBLP:conf/sosp/JiaPTWZA19}.  The compiler
community has addressed one challenge of the optimization problem,
namely the separation of the {\em algorithm} from the {\em schedule}.
This idea was introduced by the Halide language, which was
designed for high-performance code generation for image processing
pipelines~\cite{halide,DBLP:journals/cacm/Ragan-KelleyASB18}.  The
programmer writes the algorithm in an imperative, high-level language,
and writes separately a {\em schedule}, which specifies low level
optimizations, such as tiling, vectorization, or loop unrolling.
TVM~\cite{DBLP:conf/osdi/ChenMJZYSCWHCGK18} extends this principle
from image processing to tensor processing for general-purpose ML
applications.




In this work we are not concerned with schedules, but with a different
challenge in tensor processing: optimizing the query plan based on how
the tensors are stored in memory. {\em Storage} refers in this
  paper to {\em memory layout}, and not to persistent representation.
When a tensor is sparse, the programmer has many choices for
representing it in main memory, and the best plan for a tensor program
varies dramatically depending on what storage was chosen, the
statistics of the data (e.g. how sparse or dense), and the particular
tensor program.  For example, a vector $a(i)$ can be stored as a dense
array, or as a hash table indexed by $i$, or as two parallel arrays
storing the indices $i$ and the values $a(i)$.  When the tensors are
dense, then the best plan may be to use a linear algebra
library~\cite{DBLP:journals/cse/StonebrakerBZB13,DBLP:journals/cacm/LuoGGPJJ20},
while for sparse tensors a better plan may be to use relational query
operators, e.g.  hash joins.  Tensors can be easily represented as
relations, and tensor programs can be expressed as SQL
queries~\cite{viktor-leis-cidr-2022}, but relational engines are not
designed to support storage formats specifically optimized for sparse
tensors (e.g. the CSR format discussed later).


To address the storage problem, Taco~\cite{taco} separates the {\em
  tensor storage format} from the tensor program.  In the storage
format the user can specify separately for each dimension whether it
is dense or sparse, and can also order the dimensions, leading to
$d\verb+!+ \cdot 2^d$ possible formats for a $d$-dimensional
tensor.\footnote{Taco was later extended to support 6 formats per
  dimension~\cite{taco-formats}.}  Given a tensor program, the Taco
compiler generates code that optimizes the access to the storage
formats.  Taco does not perform cost-based optimizations, which means
that the programmer still needs to be aware of the storage
specification.  For example, if the vector $a$ is sparse while $b,c$
are dense, then $a(i)*(b(i)+c(i))$ is best rewritten as
$a(i)*b(i)+a(i)*c(i)$, because computing $b(i)+c(i)$ results in
  a large, dense vector, while $a(i)*b(i)$ is a small, sparse vector,
  no larger than $a$, and similarly for $a(i)*c(i)$. However, the task
  of rewriting the expression is left to the programmer.

In this paper we propose a rule-base approach to optimizing tensor
programs over flexible tensor storage, using a cost-based optimizer.
The main novelty in our approach is that the storage descriptors
themselves are also defined in the same declarative language as the
tensor program.  To specify how a tensor is stored, the user writes a
{\em storage mapping} from the physical data structures (arrays and/or
hash tables) to the logical tensor.  Our system evaluates the tensor
program by first composing it with the storage mappings, then
optimizing it using rewrite rules.  This improves in two ways over
previous systems.  First, the storage formats are no longer hard
coded, but the user is free to define their own.  For example, users
may describe one of the popular storage formats COO, CSR, etc, or
define a format optimized for upper-triangular matrices, or for
band-matrices, or a space-filling curve, etc.  There is no bound on
the number of storage representations, the only limit is the
expressivity of the query language and the power of the optimizer.
Second, the optimizer is now able to perform a rich collection of
high-level optimizations, such as factorization, loop fusion, or join
reordering, and optimize the tensor program specifically for the given
storage.  For example, the optimizer may consider both expressions
$a(i)*(b(i)+c(i))$ and $a(i)*b(i)+a(i)*c(i)$, and choose the optimal
one based on their physical storage and data statistics. To the best
of our knowledge, our system, called \system, is the first
cost-based optimizer for a declarative tensor language.  We show in
Sec.~\ref{sec:experiments} that, due to the rewrite rules, \system\
significantly outperforms both Taco~\cite{taco} (a tensor algebra) and
\duckdb~\cite{DBLP:conf/sigmod/RaasveldtM19} (an optimized relational
engine) for several tensor programs, although their physical execution
engines are as good as, or even better than ours.

\begin{figure*}
  \begin{center}
    \begin{minipage}{.52\textwidth}
      \begin{center}
        \begin{lstlisting}[linewidth=\textwidth,frame=single]
CREATE TENSOR A AS
  sum(<(i,k,l), B_v> in B, <(k,j), C_v> in C, <(j,l), D_v> in D)
    { (i, j) -> B_v * C_v * D_v }
\end{lstlisting}

        (a) MTTKRP kernel $A(i,j) = \sum_{k,l} B(i,k,l) \cdot C(k,j) \cdot D(l,j)$ in \lang.
      \end{center}

      \begin{center}
        \begin{tabular}{| c | c | c | c |}
          \multicolumn{4}{c}{\texttt{C:}} \\ \hline
          6 & 0 & 9 & 8                   \\ \hline
          0 & 0 & 0 & 0                   \\ \hline
          5 & 0 & 0 & 7                   \\ \hline
        \end{tabular}~
        \hspace{0.1\textwidth}
        \begin{tabular}{l | c | c | c | c | c |} \cline{2-2}
          \texttt{C\_len1:} & 3                                                            \\ 
          \cline{2-5}
          \texttt{C\_pos2:} & 0 & 3 & 3                         & 5 & \multicolumn{1}{c}{} \\ \cline{2-6}
          \texttt{C\_idx2:} & 0 & 2 & \multicolumn{1}{ c ||}{3} & 0 & 3                    \\\cline{2-6}
          \texttt{C\_val:}  & 6 & 9 & \multicolumn{1}{ c ||}{8} & 5 & 7                    \\ \cline{2-6}
        \end{tabular}

        (b) Matrix $C$ and its CSR format from~\cite[Fig.5(f)]{taco}.
      \end{center}

      \begin{center}
        \begin{lstlisting}[linewidth=\textwidth,frame=single]
CREATE TENSOR C AS sum(<row,_> in 0:C_len1)
    { row ->
      sum(<off,col> in C_idx2( C_pos2(row):C_pos2(row+1) ))
        { col -> C_val(off) }
    }
\end{lstlisting}

        (c) The \lang\ storage mapping for CSR.
      \end{center}
    \end{minipage}%
    \hspace{0.01\textwidth}
    \begin{minipage}{0.465\textwidth}
      \begin{center}
        \begin{lstlisting}[linewidth=\textwidth,frame=single]
sum(<i_pos, i> in B_idx1)
  { i ->
    sum(<k_pos, k> in 
          B_idx2( B_pos2(i_pos):B_pos2(i_pos+1) ))
      sum(<j_pos, j> in 
            C_idx( C_pos(k):C_pos(k+1) ))
        { 
          j ->
          C_val(j_pos) * (
            merge(<l_posB,l_posD,l> in
                <B_idx3( B_pos3(k_pos):B_pos3(k_pos+1) ),
                  D_idx( D_pos(j):D_pos(j+1) )>)
              B_val(l_posB) * D_val(l_posD)
          )
        }
  }
\end{lstlisting}

        (d) Optimized MTTKRP in \lang.
      \end{center}
    \end{minipage}%
  \end{center}

  \caption{Illustration of \system.  (a) MTTKRP kernel, (b) CSR memory
    layout, (c) CSR storage mapping, (d) optimized plan.}
  \label{fig:one}
\end{figure*}

Fig.~\ref{fig:one} illustrates how \system\ processes the matricized
tensor times Khatri-Rao product, MTTKRP~\cite{taco},
$A(i,j) = \sum_{k,l} B(i,k,l) \cdot C(k,j) \cdot D(l,j)$.
Fig.~\ref{fig:one} (a) shows the tensor program written in our
declarative tensor language \lang\ (described in
Sec.~\ref{sec:system}), while (b) shows the Compressed Sparse Row
(CSR) memory layout of matrix !C!, which is one of several formats described
in~\cite{taco,taco-formats} (reviewed in Sec.~\ref{sec:background}).
For each matrix or tensor, the user describes its memory layout by
writing a storage mapping, also in \lang; the storage mapping for the
matrix !C!  is shown in Fig.~\ref{fig:one} (c), and similar storage
mappings need to be defined for !B! and !D!.  To execute the program,
the system composes the tensor program with the storage mappings, then
chooses an optimal plan using a cost-based optimizer; the optimal plan
is shown in Fig.~\ref{fig:one} (d).  While the plan could be further
optimized for some sophisticated schedule (as done by Halide, TVM, and
Taco), we currently do not support schedules and simply run the
optimal plan directly in Julia.

The main challenge in developing the cost-based optimizer is the right
choice of tensor processing language.  All query optimizers
use the relational algebra as intermediate language.
  However, we found that a calculus, rather than an algebra, is
  better suited for optimizing tensors; here {\em calculus} refers to
  a language with explicit use of variables, while {\em algebra}
  refers to a variable-free language.  There are two reasons for that.
  First, the physical plan of a tensor program consists of for-loops
  with explicit variables.  They look like this:
%
%
!for i=1:m do for j=1:n do ...!
  instead of this: $A \Join (B \Join \cdots)$, and optimizing
  directly expressions with variables simplifies the generation of the
  physical plan.  Second, the intermediate language for tensor
  programs needs to support nested collections, which occur in sparse
  formats like CSR, CSC, CSF, while standard relational algebra, as
  well as some recent extensions to linear
  algebra~\cite{DBLP:conf/sigmod/HutchisonHS17} support only flat
  collections.  Algebras for nested collections exists, but they tend
  to be much harder to read than calculus, making it harder to design
  and debug optimization rules, e.g. compare the rules in Table
  2~\cite{DBLP:journals/pvldb/YuanJZTBJ21} to those in
  Fig.~\ref{fig:opt_rules} in our paper.  For these reasons, we opted
  for a calculus-based intermediate language.  We are not the first to
  use a calculus-based intermediate language for query optimization:
  Worst Case Optimal Join algorithms are also described as nested
  loops, in effect using a calculus as intermediate
  language~\cite{DBLP:conf/icdt/Veldhuizen14,DBLP:journals/sigmod/NgoRR13,DBLP:journals/pvldb/FreitagBSKN20}.

  We describe in this paper a language, called \lang, used both
  for writing tensor programs, and for performing optimizations.
  \lang\ has a syntax that is reminiscent of Unions of Conjunctive
Queries, but where $\wedge, \vee, \exists$ are replaced by
!*, +, sum!,
to which we add !let!-bindings, and nested {\em dictionaries} as data
model; our dictionaries are similar to those in SDQL
(Semiring-Dictionary Query
Language)~\cite{DBLP:journals/corr/abs-2103-06376}, hence we call our
language $\lang$.  Our language can express tensor programs in a
notation close to mathematics, and can express complex storage
mappings corresponding to sophisticated tensor memory layouts,
including those described in~\cite{taco,taco-formats}.  Any $\lang$
query can easily be converted directly to a physical, nested for-loop
plan, because each quantified variable $i, j, \ldots$ becomes directly
a !for!  loop over that variable.  However, it is more difficult to
design an optimizer.  For example, Selinger's dynamic programming
algorithm for join
re-ordering~\cite{DBLP:journals/computer/AstrahanBCGKLLMPPSSSSTTWY79,DBLP:conf/vldb/MoerkotteN06}
no longer applies, because in a calculus there is no explicit binary
join.  Instead, our system is entirely rule-based, and the rules must be designed for a calculus rather
than an algebra.  We designed a suite of 44 \lang-rewrite rules, and
use the equality saturation system
Egg~\cite{DBLP:journals/pacmpl/WillseyNWFTP21} as rewrite engine.  Egg
uses a compact data structure called an {\em e-graph} to represent a
large number of equivalent expressions as a graph.  However, like most
term rewriting systems, Egg does not understand variables in rules.
For our optimizer, we developed a variant of the De Bruijn index that
removes the need for explicit variable representation.

  One major motivation for our work is that most of existing
  tensor and linear algebra systems in the compilers and HPC
  communities focus on dense data; in contrast, our focus in this work
  is on sparse data.  The reason for the traditional focus on dense
  data is that Linear Algebra packages were originally developed for
  use in Physics and Engineering, where tensors are dense, and they
  support highly optimized kernels for specific operations on dense
  data.  Support in these packages for sparse data is
  rare.\footnote{Cf. GitHub issues \#43497 for TensorFlow, \#72065 for
    PyTorch, \#4332 for TVM.} TACO~\cite{taco} was the first
  recognized the need to optimize tensor programs over sparse data;
  our work falls into the latter category.

In summary, we make the following contributions in this paper:

\begin{itemize}
\item We describe the architecture of \system, where tensor programs
  and tensor storage mappings are defined in a common language, and
  optimized jointly (Sec.~\ref{sec:system}).
\item We describe a declarative tensor calculus, $\lang$, for both
  tensor programs and storage mappings, and show that it can express a
  rich variety of previously proposed storage formats, and beyond
  (Sec~\ref{sec:sd}).
\item We describe a cost-based optimizer for the tensor calculus,
  which supports a rich suite of optimizations, such as factorization,
  loop fusion, and join reordering (Sec.~\ref{sec:optimizer}).
\item Finally, we conduct an experimental evaluation showing that
  \system\ can significantly outperform other tensor processing
  systems, by using a cost-based optimizer to choose the best plan for
  the given storage representation (Sec.~\ref{sec:experiments}).
\end{itemize}

\section{Background}

\label{sec:background}

{\bf Tensors} Given a number $n$, we denote by
$\rangen{n} \defeq \set{0,1,2, \ldots, n-1}$.  Let $d \geq 1$ and let
$n_1, n_2, \ldots, n_d$ be natural numbers.  A {\em tensor with $d$
  dimensions}, is element
$\bm A \in \R^{\rangen{n_1}\times \cdots \times \rangen{n_d}}$.  A
scalar, a vector, and a matrix are tensors with 0, 1, and 2 dimensions
respectively.  Given $d$ indices,
$i_1 \in \rangen{n_1}, \ldots, i_d \in \rangen{n_d}$, we denote by
$\bm A(i_1, \ldots, i_d)$ the value of the tensor at those positions;
we call each $i_j$ a {\em dimension}.

{\bf Tensor formats} We briefly review some popular tensor formats
following~\cite{taco,taco-formats}.  A {\em dense} representation of a
tensor consists of a memory array with $n_1 n_2 \cdots n_d$ elements.
The {\em coordinate format}, COO, stores only the non-zero elements in
an array, and their coordinates in $d$ separate arrays.  For example,
the dense and COO representations of the vector $\bm v = (9, 0, 7, 5)$
are:
\newline
\begin{tabular}{l | l}
  DENSE:  &
  COO: \\[-0.2em]
\begin{tabular}{l | c | c | c | c |}\cline{2-2}
  \texttt{v\_len:} & 4 \\ \cline{2-5}
  \texttt{v\_val:} & 9&0&7&5 \\ \cline{2-5}
\end{tabular}
\hspace{1em}
&
\hspace{1em}
\begin{tabular}{c | c | c | c | } \cline{2-3}
  \texttt{v\_pos:} &  0 & 3 \\ \cline{2-4}
  \texttt{v\_idx:}   &  0 & 2 & 3 \\ \cline{2-4}
  \texttt{v\_val:} &   9&7&5 \\ \cline{2-4}
\end{tabular} \\
\end{tabular}
\newline
To access $v(i)$ using the COO representation one has to first find
$i$ in $\texttt{v\_idx}$, in other words one has to find $p$ such that
$\texttt{v\_idx}(p)=i$, then return $\texttt{v\_val}(p)$; the role of
$\texttt{v\_pos}$ will become clear shortly.  The COO representation
of a matrix has two index arrays,
$\texttt{v\_idx1}$, $\texttt{v\_idx2}$, storing the rows and columns of
the non-zero element respectively.  The COO representation is compact,
but no longer enables constant-time lookup.  A {\em hash-map}
representation of the matrix is a hash-map where the keys are pairs
$(i,j)$.  It is compact and allows access in time $O(1)$, but no
longer supports a scan in either row-major or column-major order.

The Taco system~\cite{taco} describes a general scheme for storage
formats where the user can choose an order of the $d$ dimensions, and
specify, independently for each dimension, whether it is dense or
sparse.  This allows for $d\verb+!+\cdot 2^d$ formats.  The storage
uses {\em segmented} arrays, which consist of the concatenation of
several sub-arrays stored in a single array, with their starting
positions stored in a separate array.  For example, the {\em
  sparse-sparse} representation of the matrix $\bm C$ in
Fig.~\ref{fig:one} (b) is the following (taken
from~\cite[Fig.5(g)]{taco}):
\newline
\begin{tabular}{l | c | c | c | c | c |} \cline{2-3}
  \texttt{C\_pos1:} & 0 & 2 \\ \cline{2-3}
  \texttt{C\_idx1:} & 0 & 2 \\ \cline{2-4}
  \texttt{C\_pos2:} & 0 & 3 & 5 \\ \cline{2-6}
  \texttt{C\_idx2:} & 0 & 2 & \multicolumn{1}{ c ||}{3} & 0 & 3 \\ \cline{2-6}
  \texttt{C\_val:} & 6 & 9 & \multicolumn{1}{ c ||}{8} & 5 & 7 \\ \cline{2-6}
\end{tabular}
\newline
The arrays $\texttt{C\_idx2}$ and $\texttt{C\_val}$ contain two segments each: the first
segment represents row $0$ of the matrix $\bm C$, $(6,0,9,8)$; the second segment represents
row $2$, $(5,0,0,7)$.  The segments are delimited by $\texttt{C\_pos2}$, which indicates
their starting point. The row number of each segment is stored in $\texttt{C\_idx1}$: only
the values $i=0$ and $i=2$ occur here because row 1 is empty.  Alternatively, the {\em
dense-sparse} representation, shown in Fig.~\ref{fig:one} (b) stores {\em every} row,
including row 1, and for that reason there is no need to store the vector $\texttt{C\_idx1}$
(since this vector would be $(0,1,2)$), but we only store its length, $\texttt{C\_len1}=3$.
The dense-sparse representation is called {\em compressed sparse row}, or CSR, and the
sparse-sparse representation is called \emph{doubly CSR}, or DCSR.  In a later
reference~\cite{taco-formats} the authors extended the number of choices available at each
dimension from 2 to 6.

{\bf Tensors as Relations} Any $d$-dimensional tensor can be
represented as a relation with $d+1$ attributes.  For example, a
matrix $A$ can be represented as a relation $R(i,j,v)$, where $i,j$ is
the primary key, and $v$ the value of $A(i,j)$.  A clustered index on
$(i,j,v)$ corresponds roughly to a row-major ordering of the matrix; a
hash-index corresponds to a hash-map representation; while a
column-oriented storage~\cite{DBLP:journals/ftdb/AbadiBHIM13}
corresponds to a COO representation.  However, since relations are
unordered, it is not possible to use some of the other formats, like
CSR or CSC.

{\bf Semiring Dictionary} A {\em semiring} is a quintuple
$(S,+,*,0,1)$, where $S$ is a set, the operations $+,*$ are
associative with identities 0
and 1 respectively, $+$ is commutative, $*$ distributes over $+$, and $0*x=x*0=0$.  For
example, the real numbers form a semi-ring, $(\R,+,*,0,1)$. A {\em
  semiring dictionary}, or simply a {\em dictionary}, is a mapping
$K \rightarrow S$, from a finite set of keys $K$ to values in some
semiring $S$~\cite{DBLP:journals/corr/abs-2103-06376}.
If $k_1, \ldots, k_m$ are distinct keys, then
$\set{k_1 \rightarrow v_1, \ldots, k_m \rightarrow v_m}$ denotes the
dictionary that maps each key $k_i$ to the value $v_i$, and maps each
key $k \not\in \set{k_1, \ldots, k_m}$ to $0$.  In other words,
missing keys default to 0; it follows that
$\set{k_1 \rightarrow 0, k_2 \rightarrow 0, \ldots}=\set{\ }$, in
other words a dictionary containing only 0 values is the same as empty
dictionary.  In this paper the key space is always of the form
$K \defeq \rangen{n_1} \times \cdots \times \rangen{n_d}$, and we view
interchangeably a tensor $\bm A \in \R^K$ as a dictionary
$\bm A : K \rightarrow \R$.  When $d=0$, then the dictionary is of the
form $\set{() \rightarrow v}$, which we identify, with some abuse,
with the scalar value $v$.

It was observed in~\cite{DBLP:journals/corr/abs-2103-06376} that semiring
dictionaries generalize K-relations~\cite{green2007provenance}; the
set of semiring dictionaries over a fixed key-set $K$ forms another
semiring, where the plus and multiplication are defined element-wise.  For
example, if $\bm A, \bm B$ are two $m \times n$ matrices, then they
both can be viewed as dictionaries
$\rangen{m} \times \rangen{n} \rightarrow \R$, and $\bm A+\bm B$,
$\bm A*\bm B$ denote their element-wise sum and product respectively.
One consequence is that we obtain the following rule:
\begin{align*}
 \set{k_1 \rightarrow v_1} + \set{k_2 \rightarrow v_2} =
&
  \begin{cases}
    \set{k \rightarrow v_1+v_2} & \mbox{if $k_1=k_2=k$} \\
    \set{k_1 \rightarrow v_1, k_2 \rightarrow v_2} & \mbox{if
      $k_1 \neq k_2$}
  \end{cases}
\end{align*}
Another consequence is that one can define nested dictionaries, by
defining a dictionary whose values are other dictionaries.  For
example, let $S \defeq [\rangen{n} \rightarrow \R]$ denote the set of
dictionaries with keys $\rangen{n}$ and real values.  A dictionary in
$S$ is a vector of length $n$.  Then, a dictionary $\bm A: \rangen{m}
\rightarrow S$ is a vector of length $m$ of vectors of length $n$,
which is equivalent to a matrix.

\lstset{language=sdql}

{\bf SDQL} We briefly review here SDQL
from~\cite{DBLP:journals/corr/abs-2103-06376}.  In a nutshell, the
query language SDQL is like Unions of Conjunctive Queries,
where $\exists, \wedge, \vee$ are replaced with !sum!, $*$, $+$, and
the head variables are moved to the end of the query expression.  We
show here side-by-side in CQ and SDQL how to transform a vector $V$ by
removing its negative values (equivalently: setting them to 0) and
multiplying the others by 5:
\newline
\begin{tabular}{l  |  l}
CQ: & SDQL: \\
  $Q(i,5*v) := V(i,v) \wedge (v>0)$
&
\begin{minipage}{0.6\linewidth}
\begin{lstlisting}
sum( <i,v> in V )
   if v>0 then { i -> 5*v }
\end{lstlisting}
\end{minipage}
\end{tabular}
\newline
The semantics of SDQL uses the fact that dictionaries form a semiring,
i.e. can be ``added''.  The SDQL query above is executed by iterating
over the pairs $<i,v>$ in $V$, and summing up singleton dictionaries.
For example, if the vector $V$ is $(v_0, v_1, v_2, v_3, v_4)$, where
$v_0, v_3, v_4 > 0$ and $v_1, v_2 < 0$, then the query above returns
$\{0 \rightarrow 5v_0\} + \{3 \rightarrow 5v_3\} + \{4 \rightarrow
5v_4\}=\set{0 \rightarrow 5v_0, 3\rightarrow 5v_3, 4 \rightarrow
  5v_4}$.  For another illustration, the following two SDQL queries
compute the dot product $\sum_i u_i v_i$ and the element-wise product
$(u_i v_i)_i$ of two vectors $U,V$ respectively:

\begin{lstlisting}
  sum(<i,u> in U, <i,v> in V) {() -> u*v}
  sum(<i,u> in U, <i,v> in V) {i -> u*v}
\end{lstlisting}

The operator \code{*} is overloaded to define the multiplication of scalars and dictionaries. For example, !a * V! represents a scalar-vector multiplication and is equivalent to the following SDQL query:

\begin{lstlisting}
  sum(<i,v> in V) {i -> a * v}
\end{lstlisting}


\section{\system}

\label{sec:system}

\begin{figure}
  \centering
\includegraphics[width=\columnwidth]{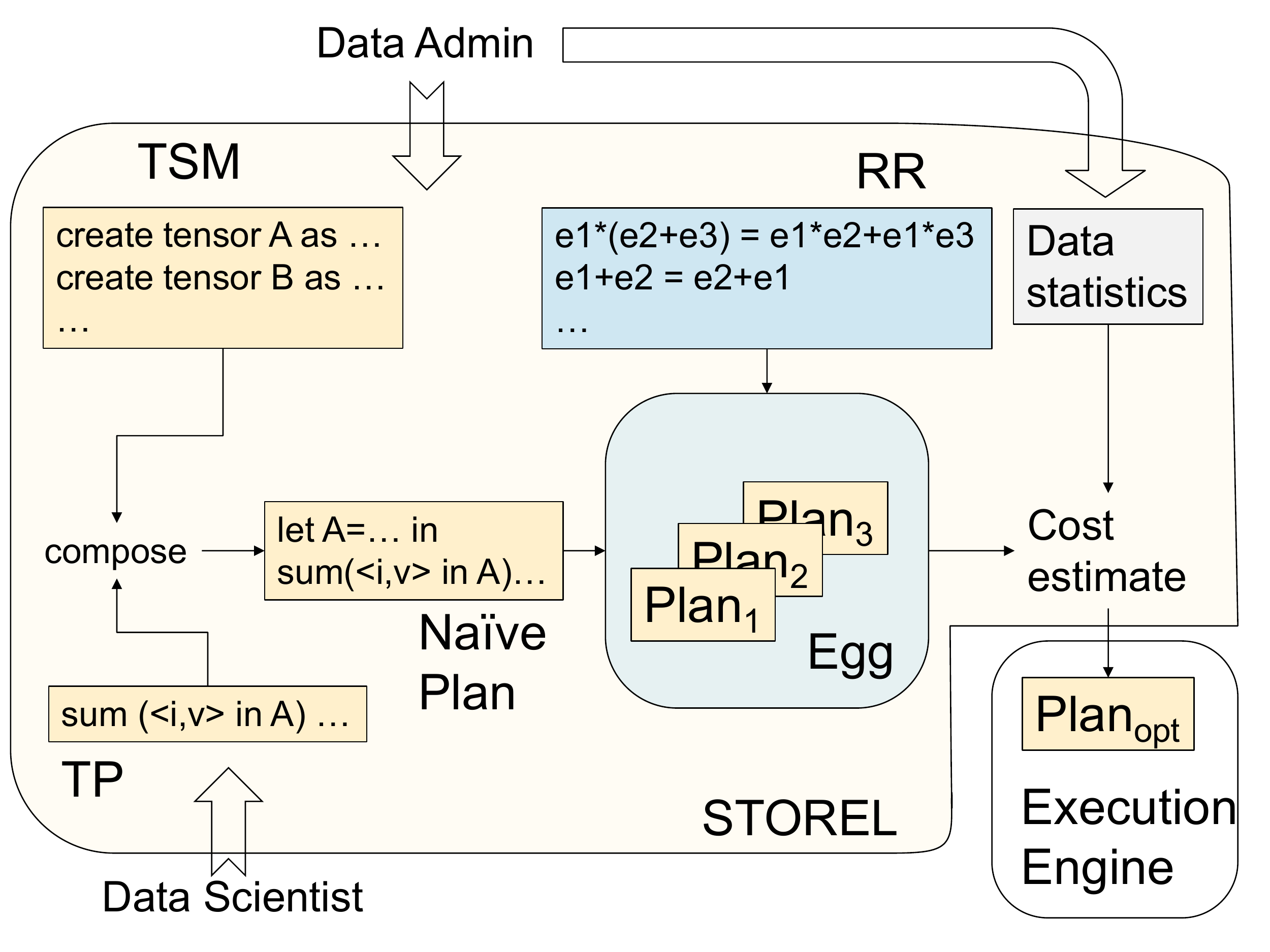}
\vspace{-0.8cm}
\caption{System's architecture.  TSM$=$Tensor Storage Mapping;
  TP$=$Tensor Program; RR$=$Rewrite Rules.}
  \label{fig:arch}
  \vspace{-0.4cm}
\end{figure}

Here we describe architecture of our system \system, and its
declarative language \lang.

\subsection{Architecture}

The architecture of \system\ is shown in Fig.~\ref{fig:arch}.  All
yellow-gold boxes represent \lang\ programs, described below, while
the blue box represents rewrite rules described in
Sec.~\ref{sec:optimizer}.  The end user (for example a data scientist)
writes a Tensor Program (TP) in \lang.  Separately, the data
administrator (possibly the same user) familiar with low level
optimizations, writes Tensor Storage Mappings (TSM), one for each
tensor.  \system\ composes the two expressions, by substituting each
tensor variable in TP with its corresponding definition in TSM.  This
results in a \lang\ expression which we call the Naive Plan.  The
Naive Plan is then submitted to the Egg equality saturation
system~\cite{DBLP:journals/pacmpl/WillseyNWFTP21}.  Egg has access to
a knowledge base of Rewrite Rules (RR), and applies all rules until
saturation, i.e. until no more rule can be applied.  The current
collection of rewrite rules includes about 44 rules, described in
Sec.~\ref{sec:optimizer}.  Egg stores all equivalent plans in a very
compact data structure called an {\em e-graph}.  Next, \system\ uses
data statistics and a simple cost model to associate a cost to each
equivalent plan; currently, the user needs to provide the data
statistics manually.  Egg then extracts the cheapest plan from the
e-graph, and this plan is finally submitted to the execution engine.
We currently use Julia~\cite{julia} as our execution
engine. Alternatively, the optimal plan could be further optimized by
applying {\em schedules}, but this is not currently supported in our
system.



\subsection{\lang}

\label{subsec:sdqlite}

  \system\ needs a language to express the tensor programs, a
  formalism for expressing tensor storage formats, an intermediate
  language in which to express the optimizations, and a physical
  language in which the programs are executed.  In this paper we are
  introducing a declarative language called \lang, which serves the
  first three purposes: it can express Tensor Programs (TP), it can
  express sophisticated Tensor Storage Mappings (TSM), and we also use
  it as intermediate language for performing optimizations.  \lang\
  can be easily converted to physical plans, as we describe in
  Sec.~\ref{sec:optimizer}.  A language that satisfies all these goals
  requires a careful design: we describe \lang\ in this section, and
  note that it is derived from
  SDQL~\cite{DBLP:journals/corr/abs-2103-06376}; we discuss the
  differences at the end of this section.


The data model for \lang\ consists of {\em scalars} (integers or
reals), and {\em nested dictionaries}.  The latter have type
$\rangen{n} \rightarrow S$, where the value space $S$ is the set of
integers, reals, or another dictionary.  Thus, the data
model in \lang\ consists of the following types:
\begin{align*}
& \rangen{n_1} \rightarrow \rangen{n_2} \rightarrow \cdots \rangen{n_d} \rightarrow \R\\
& \rangen{n_1} \rightarrow \rangen{n_2} \rightarrow \cdots \rangen{n_d} \rightarrow \Z
\end{align*}
where $d \geq 0$; when $d=0$ then these are scalar types.  One can
equivalently view the data model of \lang\ as consisting of
real- and integer-valued tensors.  To see this, we recall the
{\em curry} operation, which converts a function of type
$K \times K' \rightarrow S$ into a function of type
$K \rightarrow [K' \rightarrow S]$, and the {\em uncurry} operation
which goes the other way.  By repeatedly uncurrying a nested
dictionary, we can convert it to a
$n_1 \times n_2 \times \cdots \times n_d$ tensor.  For that reason, in
this paper we blur the distinction between (nested) dictionaries and
tensors, and view the data model of \lang\ as consisting of tensors.

{\bf Syntax} The expressions in \lang\ are the following:

\begin{lstlisting}
  e1+e2,    e1*e2,    {k -> e},    e(i), i1:i2, e(i1:i2),
  if (c) then e,  let v = e1 in e2,   sum(<k,v> in e1) e2
\end{lstlisting}
where !k!, !v! are variables, !e!, !e1!, !e2!, !e3! are dictionary
expressions, !i!, !i1!, !i2! are index expression, i.e.  of type
!int!, and !c! is a Boolean expression.  Each of the \lang\
expressions above returns a dictionary, which may, in particular, be a
scalar.  We also include standard primitive operations over scalars of
type real or integer, like
division !/!,
modulo !
exponentiation !exp(...)!
comparison !e1<e2!, etc.

{\bf Semantics} We briefly describe the semantics of \lang.  !e1+e2!
and !e1*e2! compute the sum and product of dictionaries; !e(i)!
applies the dictionary !e! to the key !i!.

The range expression !i1:i2! returns the dictionary consisting of the
sum of !{i -> i}!, for all !i! in the range !i1,...,i2-1!, i.e.:
\begin{lstlisting}
  i1:i2 = { i1 -> i1, i1+1 -> i1+1, ..., i2-1 -> i2-1 }
\end{lstlisting}
The sub-array expression !e(i1:i2)! is used for representing segmented arrays and returns the following dictionary:
\begin{lstlisting}
  e(i1:i2) = { i1->e(i1),i1+1->e(i1+1),...,i2-1->e(i2-1) }
\end{lstlisting}
The conditional
!if (c) then e!
returns !e! if !c! is true, and returns zero (!0! or !{}! depending on the type of !e!) otherwise, and the !let!
construct introduces a temporary variable !v!, with value !e1!, which
may be used in !e2!.

We explain the summation.  Assume the value of !e1!  is:
\begin{lstlisting}
e1 = { k1 -> v1, ..., kn -> vn }
\end{lstlisting}
Then the value of
!sum(<k,v> in e1) e2!
is:
\begin{lstlisting}
  e2[k1/k,v1/v] + e2[k2/k,v2/v] + ... + e2[kn/k,vn/v]
\end{lstlisting}
where !e2[k_i/k,v_i/v]! represents the result of substituting in !e2!
the variables !k!, !v!  with the values !k_i! and !v_i!.

{\bf Syntactic Sugar} We also included some convenient syntactic sugar
extensions in \lang, described in Table~\ref{tbl:desugar}.

\begin{table}
\begin{center}
\begin{tabular}{|l|l|l|} \hline
Construct & Desugars to & Notes \\ \hline
!e(e1,e2)! & !e(e1)(e2)! & curry \\ \hline
!{ (e1, e2) -> e }! & !{ e1 -> { e2 -> e } }! & curry \\ \hline
\begin{lstlisting}
sum(<(k1,k2),v> in e1)
  e2
\end{lstlisting}
            &
\begin{lstlisting}
sum(<k1,w> in e1)
  sum(<k2,v> in w) e2
\end{lstlisting}
 &uncurry \\ \hline
\begin{lstlisting}
let v1=e1, v2=e2
in ...
\end{lstlisting}
&
\begin{lstlisting}
  let v1=e1
  in let v2=e2
     in ...
\end{lstlisting}
 & \\ \hline
\begin{lstlisting}
sum(<k1,v1> in e1,
    <k2,v2> in e2) e3
\end{lstlisting}
&
\begin{lstlisting}
sum(<k1,v1> in e1)
 sum(<k2,v2> in e2)
  e3
\end{lstlisting}
&  \\ \hline
\begin{lstlisting}
sum(<k,v1> in e1,
    <k,v2> in e2) e3
\end{lstlisting}
&
\begin{lstlisting}
sum(<k1,v1> in e1)
 sum(<k2,v2> in e2)
  if (k1 == k2) then e3
\end{lstlisting}
& \begin{tabular}{@{}l@{}}!k! is used\\ twice in\\ LHS\end{tabular}  \\ \hline
\end{tabular}
\end{center}
\caption{Syntatic sugar extensions in \lang.}
\label{tbl:desugar}
\vspace*{-2em}
\end{table}


\begin{example} \label{ex:matrix:multiplication}
For a simple illustration, the following \lang\ query computes the
product of two matrices !A! and !B!:
\begin{lstlisting}
  sum( <(i,j),a> in A, <(j,k),b> in B) { (i,k) -> a*b }
\end{lstlisting}
It is internally desugared to:
\begin{lstlisting}
  sum( <i,rowA> in A) sum(<j1,a> in rowA)
      sum(<j2,rowB> in B) sum(<k,b> in rowB)
          if (j1==j2) { i -> { k -> a*b }}
\end{lstlisting}
Notice the power of viewing dictionaries as semirings.  The semantics
of query above consists of emitting  $n^3$ singleton dictionaries of
the form
!{i->{k-a*b}}!,
which are ``added'' up and result in only $n^2$ pairs !i,k!.
``Addition'' acts like a group-by, in other words we have:
\newline
\begin{lstlisting}
  {i -> {k -> ai1*b1k}} + {i->{k->ai2*b2k}} + ...
= {i -> {k -> ai1*b1k} + {k->ai2*b2k} + ... }
= {i -> {k -> ai1*b1k+ai2*b2k+...}}
\end{lstlisting}

Alternatively, if the dimensions of the two matrices are known to be
$m \times n$ and $n \times p$, then matrix multiplication can be
written as:
\begin{lstlisting}
  sum( <i,_> in 0:m, <j,_> in 0:n, <k,_> in 0:p)
     { (i,k) ->  A(i,j)*B(j,k) }
\end{lstlisting}
\end{example}

{\bf Discussion} \lang\ is a declarative language, in that it
  does not specify the order of operations.  This similar to, say,
  SQL, where the order of the tables in the !FROM!, or the
  order of the predicates in the !WHERE! clause do not specify
  that the joins, or the evaluation of predicates, need to be executed
  in that order.  In fact, \lang\ is basically UCQ, where
  $\exists, \wedge, \vee$ are replaced with !sum!, $*$, $+$, as
  we explained in Sec.~\ref{sec:background}; UCQ is generally accepted
  to be a declarative language, and \lang\ is similarly declarative.
While \lang\ borrows several ideas from
SDQL~\cite{DBLP:journals/corr/abs-2103-06376}, it differs in some
important ways, as follows.  \lang\ adds subarray expressions
  (needed in Sec.~\ref{sec:sd}), the \code{merge} operator (needed in
  Sec.~\ref{sec:physical}), and has several syntactic sugar
  extensions, shown in Table~\ref{tbl:desugar}.  We restricted the
  keys to be numbers only (while SDQL allows records and dictionaries), which is
  necessary to enable the optimization rules in
  Sec.~\ref{sec:rewrite}.  Finally, we defined cardinality and cost
  estimation rules and extended optimization rules, as described in
  Sec.~\ref{sec:optimizer}. 

\section{Tensor Storage Mappings}

\label{sec:sd}

We have described in the previous section the declarative tensor
language \lang, which can be used to write tensor programs in a
notation close to a mathematical notation; as we discussed, we use
tensors and dictionaries interchangeably in this paper.  So far, all
tensors manipulated in \lang\ have only a logical data model, and we
have not defined yet a physical model.  In this section we extend
\lang\ with a physical data model, and show how we use it to define
Tensor Storage Mappings (TSM), from a physical to a logical
representation.

The physical data model in \lang\ consists of four data types: (a)
scalar values, which can be of type !real! or !int!, (b) arrays of
type !int! or !real!, and (c) hash-maps that map tuples of integers to
!int! or !real!, or (d) tries, which are trees of hash-maps.  The
data admin (Fig.~\ref{fig:arch}) defines named data values
using the following syntax:
\begin{lstlisting}
 CREATE [real | int] SCALAR U;
 CREATE [real | int] ARRAY A(n);
 CREATE [real | int] HASHMAP B(n1, n2, ..., nd);
 CREATE [real | int] TRIE C(n1)(n2)...(nd);
\end{lstlisting}
Here !U!, !A!, !B!, !C! are names, i.e. identifiers, !n!, !n1!,
$\ldots$, !nd! are expressions of type !int!.  An array is a
continuous memory array of fixed size !n!.  Both a hash-map and a trie
logically represent a dictionary
$\rangen{n_1} \times \cdots \times \rangen{n_d} \rightarrow \R$ (or
$\cdots \rightarrow \Z$), but differ at the physical level.  The
hash-map maps a key $(i_1, \ldots, i_d)$ to a real or integer value,
while a trie is a hash map that maps a key $i_1$ to another hash-map
of type
$\rangen{n_2} \times \cdots \times \rangen{n_d} \rightarrow \R$.  All
symbols !U!, !A!, !B! are global symbols, in contrast to symbols
introduced by the !let!  binding, which are local.

Next, the data administrator writes a Tensor Storage Mapping (TSM) for
each logical tensor, using the following statement:
\begin{lstlisting}
  CREATE TENSOR T AS ...;
\end{lstlisting}
where the ellipsis represent a \lang\ expressions that uses the named
data values defined earlier, and returns a logical tensor (dictionary)
!T!.

By combining a simple physical data model with a powerful tensor
language, \lang\ allows the data administrator to define sophisticated
storage mappings, which helps the administrator exploit the particular
characteristics of her tensors.

We illustrate with several TSM examples, showing various formats for
representing a matrix !C!.

\begin{example} \label{ex:c:dense:matrix}
  The following TSM defines a dense, row-major representation of !C!:
\begin{lstlisting}
CREATE int SCALAR M, N; CREATE ARRAY V(M*N);
CREATE TENSOR C AS
  sum (<i,_> in 0:M, <j,_> in 0:N) { (i,j) -> V(i*N+j) };
\end{lstlisting}
\end{example}

\begin{example} \label{ex:c:sparse:sparse} Suppose we want to store
  !C! using the DCSR format (sparse-sparse) described in
  Section~\ref{sec:background}.  Then we need to define the following
  physical data types, and TSM:
\begin{lstlisting}
CREATE int ARRAY C_pos1(2);
CREATE int ARRAY C_idx1(C_pos1(1));
CREATE int ARRAY C_pos2(C_pos1(1)+1);
CREATE int ARRAY C_idx2(C_pos2(C_pos1(1)));
CREATE real ARRAY C_val(C_pos2(C_pos1(1)));
CREATE TENSOR C AS
  sum (<i_pos, i> in C_idx1)
     let j_start = C_pos2(i_pos),
         j_end   = C_pos2(i_pos+1)
     in sum( <j_pos, j> in C_idx2( j_start:j_end ))
            { (i,j) -> C_val(j_pos)}
\end{lstlisting}
We then materialize the physical data types as follows:
\begin{itemize}
\item !C_pos1! has size 2 and values !C_pos1(0)=0!, !C_pos1(1)=!  the
  number of non-empty rows in the matrix !C!.
\item !C_idx1!  contains all indexes !i! of the nonempty rows in !C!,
  in increasing order.
\item !C_pos2! defines the segments in the arrays !C_idx2! and
  !C_val!; there is one segment for each non-empty row in the matrix,
  hence the size of !C_pos2! is the number of non-empty rows plus 1.
  Its last position defines the sizes of !C_idx2! and !C_val!.
\item Finally, !C_idx2! and !C_val! contain the segmented arrays that
  represent the non-empty rows of !C! as sparse vectors.
\end{itemize}
The !TENSOR! expression defines how to build !C! from these
arrays.

For a similar example, Fig.~\ref{fig:one} (c) defines the Tensor
Storage Mapping from the CSR representation Fig.~\ref{fig:one} (b) to
the matrix !C!.
\end{example}

\begin{example} \label{ex:c:hash}
  We briefly illustrate !HASHMAP! and !TRIE!.  The following
  represents  !C! using a !HASHMAP!:
\begin{lstlisting}
CREATE real HASHMAP H(M,N);
CREATE TENSOR C AS sum(<(i,j),v> in H) {(i,j) -> v};
\end{lstlisting}
  This is commonly known as the Dictionary of Keys (DOK) format in
  SciPy~\cite{dokformat}.

  Alternatively, we could store !C! in a !TRIE! of depth 2:
\begin{lstlisting}
CREATE real TRIE T(M)(N);
CREATE TENSOR C AS
  sum (<i,row> in T, <j,v> in row) { (i,j) -> v };
\end{lstlisting}
  The difference is that now the key of the hash-map !T! is a single
  index !i!, and the value is another hash-map that maps the columns
  !j! to values.
\end{example}

We end this section by emphasizing that storage mappings defined in a
declarative language like \lang\ are significantly more expressive
than fixed, predefined storage formats.  For example, it is easy to
represent in \lang\ a storage mapping for a dense lower-triangular
matrix !A!, a band matrix !B! (where !B(i,j)<>0! only when
!abs(i-j)<=1!), or the Z-order space-filling curve !C!, although
\lang\ was not explicitly designed for these types of storages:
\begin{lstlisting}
CREATE real ARRAY A_val(N*(N+1)/2);
CREATE TENSOR A AS        // lower triangular
  sum(<i,_> in 0:N, <j,_> in 0:(i+1))
     {(i,j) -> A_val(i*(i-1)/2+j)}

CREATE real ARRAY B_val(3*N-2);
CREATE TENSOR B AS        // band matrix
  sum(<p,_> in 0:N)
    { (p,p) -> B_val(3*p)} +
    if (p<N)
      then { (p,p+1) -> B_val(3*p+1),
             (p+1,p) -> B_val(3*p+2) }

CREATE real ARRAY C_val(N*N);   // N is power of 2
CREATE TENSOR C AS              // Z-order curve
   sum (<d,v> in C_val)
      let i = even_bits(d),     // even bits of d
          j = odd_bits(d)       // odd bits of d
      in { (i,j) -> v }
\end{lstlisting}

\section{Optimizations}

\label{sec:optimizer}

We have seen that tensor processing in \system\ consists of two
separate tasks: writing the Tensor Program (TP), and writing the
Tensor Storage Mappings (TSM), see Fig.~\ref{fig:arch}.  The TP simply
refers to logical tensor names, like !A! or !B! or !C!, while the TSM
describes how these tensors are stored in physical arrays, hash-maps,
or tries.  Both programs are expressed at a {\em logical} level, in
the same declarative language \lang.  In this section we describe how
\system\ combines these two into a single, optimized {\em physical}
plan, which can be directly executed by an engine; we currently use
Julia as our physical execution engine.  In this section we will
refer to any logical \lang\ query as a {\em logical plan}.  Then, we
describe some refinements of the logical operators into physical
operators: a query expression using the physical operators will be
called {\em physical plan}.

\subsection{The Naive Logical Plan}

\lstset{language=sdql}

\begin{figure*}
\begin{tabular}{|l l c l|}
\hline
\multicolumn{4}{|l|}{\textit{Associativity/Commutativity Rules:}} \\ \hline
A1:&!e1 * (e2 * e3)! & \transbi & !(e1 * e2) * e3! \\ \hline
A2:&!{ e1 -> e2 * e3 }! & \transbi & !{ e1 -> e2 } * e3! \\ \hline
A3:&!{ e1 -> e2 * e3 }! & \transbi & !e2 * { e1 -> e3 }! \\ \hline
A4:&!if(e1) then e2 * e3! & \transbi & !e2 * if(e1) then e3! \\ \hline
C1:&!e1 + e2! & \transbi & !e2 + e1! \\ \hline
C2:&!e1 == e2! & \transbi & !e2 == e1! \\ \hline
\end{tabular}\hspace{1cm}
\begin{tabular}{|l l c l|}
\hline
\multicolumn{4}{|l|}{\textit{Algebraic Simplifications:}} \\ \hline
L1:&!e + 0! & \transto & !e! \\ \hline
L2:&!e * 0! & \transto & !0! \\ \hline
L3:&!e * 1! & \transto & !e! \\ \hline
L4:&!-0! & \transto & !0! \\ \hline
L5:&!e - 0! & \transto & !e! \\ \hline
L6:&!e - e! & \transto & !0! \\ \hline
\end{tabular}
\begin{tabular}{|l l c l|}
\hline
\multicolumn{4}{|l|}{\textit{Distributivity (Factorization) Rules:}} \\ \hline
D1:&!e1 * e2 + e1 * e3! & \transbi & !e1 * (e2 + e3)! \\ \hline
D2:&!sum(<k,v> in e1) e2 * e3! \rulecondition{} !k,v! $\notin$ $FV$(!e2!) & \transbi & !e2 * (&sum(<k,v> in e1) e3)! \\ \hline
D3:&!sum(<k,v> in e1) e2 * e3! \rulecondition{} !k,v! $\notin$ $FV$(!e3!) & \transbi & !(sum(<k,v> &in e1) e2) * e3! \\ \hline
D4:&!sum(<k,v> in e1) { e2 -> e3 }! \rulecondition{} !k,v! $\notin$ $FV$(!e2!) & \transbi & !{ e2 -> sum(<k,v> in e1) e3 }! \\ \hline
\hline
\multicolumn{4}{|l|}{\textit{Fusion Rules:}} \\ \hline
&!sum(<k,v> in e1)! & & !let k = e2 in!\\
F1:&!  if(k == e2) then! \rulecondition{} !k,v! $\notin$ $FV$(!e2!) & \transbi & !let v = e1(k) in!\\
&!    e3!   &  &
!e3!
 \\ \hline
&  !sum(<k1,v1> in !  & &
!sum(<k2,v2> in e1)!
 \\
F2:& !        (sum(<k2,v2> in e1) {k2 -> e2}))! &\transbi & !  let k1=k2, v1=e2 in! \\
& !  e3! & & !  e3!\\ \hline
&!sum(<k1,v1> in !  & &
!sum(<k2,v2> in e1)!
 \\
F3:& !        (sum(<k2,v2> in e1) {@unique e2 -> e3}))! &\transbi & !  let k1=e2, v1=e3 in! \\
& !  e4! & & !  e4!\\ \hline
&!sum(<k1,v1> in e1)! & & !merge(<k1,k2,v1> in <e1,e2>)! \\
F4:&!  sum(<k2,v2> in e2)! \rulecondition{} !k1,v1! $\notin$ $FV$(!e2!)  & \transbi & !  let v2 = v1 in! \\
&!    if(v1==v2) then e3!  & & !  e3!
 \\ \hline\hline
\multicolumn{4}{|l|}{\textit{Dictionary Rules:}} \\ \hline
T1:&!sum(<k,v> in e) { k -> v }!  & \transbi &
!e!
 \\ \hline
T2:&!e2(e1) + e3(e1)! & \transbi & !(e2 + e3)(e1)! \\ \hline
T3:&!{ e1 -> e2 } + { e1 -> e3 }! & \transbi & !{ e1 -> e2 + e3 }! \\ \hline
T4:& !(e1:e2)(e3)!  & \transbi &
!if(e3 >= e1 && e3 < e2) then e1 + e3!
 \\ \hline
T5:&!sum(<k,v> in e1:e2) e3! & \transbi &
 !sum(<k,_> in e1:e2) let v=k+e1 in e3! \\ \hline
\end{tabular}
\vspace{-0.4cm}
\caption{Selected transformation rules of the 44 rules that form the basis of our
cost-based optimizer.}
\vspace{-0.4cm}
\label{fig:opt_rules}
\end{figure*}

The first step of the optimizer consists of composing the Tensor
Program with the Tensor Storage Mappings, to obtain the {\em Naive
  Logical Plan}, obtained by simply appending the TSM and the TP.
More precisely, if the TP operates over tensors !A!, !B!, $\ldots$,
and each is defined by one TSM, then the naive logical plan looks like
this:
\begin{lstlisting}
  let A = TSM-for-A
      B = TSM-for-B
      ...
  in TP
\end{lstlisting}
The input to the Naive Logical Plan consists of the physical arrays,
hash maps, and tries mentioned in the TSMs.  Its output is the final
answer of the TP.

Evaluating the naive plan directly is very inefficient, because it
involves materializing all tensors in some naive representation, which
is what we wanted to avoid in the first place.  Instead, the system
performs a sequence of {\em logical rewritings} in order to optimize
the program.

\subsection{Logical Rewritings}
\label{sec:rewrite}

Our optimizer is a cost-based optimizer that applies a set of
  rules to find expressions equivalent to the given query, then uses a
  cost model to select the cheapest expression. It uses
Egg~\cite{DBLP:journals/pacmpl/WillseyNWFTP21} for simplifying a
\lang\ expression using the rules.  Our rule base currently consists
of 44 rules of the form:
\newline
\begin{tabular}{ccc}
  !Pattern1! & \transto & !Pattern2!
\end{tabular}
\newline
When we  assert rules in both directions then we write:
\newline
\begin{tabular}{ccc}
  !Pattern1! & \transbi & !Pattern2!
\end{tabular}

We show a few selected rules in Fig.~\ref{fig:opt_rules}.  We start by
showing some simple associativity and commutativity rules, followed by
algebraic simplifications rules, which are unidirectional.  The
factorization rules allow us to move constant factors in or out of the
summation; as we explain in Sec.~\ref{sec:experiments}, this leads to
some significant performance improvements.  The next group contains
loop fusion rules, which are known to be of key importance for linear
algebra or tensor algebra~\cite{DBLP:journals/pvldb/BoehmRHSEP18}.
Finally, the dictionary rules capture the way that summation interacts
with dictionaries.

\begin{example} For a simple illustration, we show how to convert an
  iteration into a lookup.  Consider the following inner product of
  two vectors:
\begin{lstlisting}
sum (<i,a> in A, <i,b> in B) { () -> a*b }
\end{lstlisting}
After desugaring the query becomes:
\begin{lstlisting}
sum (<i,a> in A) sum (<j,b> in B) if (i==j) { () -> a*B }
\end{lstlisting}
At this point the optimizer can apply fusion rule !F1! and rewrite the
query to:
\begin{lstlisting}
sum (<i,a> in A) let k=j, v=b(k) in a*v
\end{lstlisting}
When !A! is sparse and !B! is stored as a hash map, then this
expression is much more efficient, because it iterates only over the
non-zero elements of !A!, and uses a lookup to retrieve the values of
!B!.
\end{example}

%

{\em Unique Constraint} To increase the power of our optimizer,
we have extended \lang\ with a constraint called !@unique!, which may
be specified in a dictionary construction:
\begin{lstlisting}
  { @unique k -> e }
\end{lstlisting}
The semantics of !@unique! is that, in a !sum!, all keys are asserted
to be distinct.  Fusion rule !F3! requires the !@unique!  constraint,
and allows two nested loops to be fused into a single loop.  The role
of !@unique! is only to inform the optimizer: it has no effect at
runtime.

We explain now the rule !F3!.  Consider the following subexpression of
the LHS of the rule:
\begin{lstlisting}
  sum(<k2,v2> in e1) {@unique e2 -> e3}
\end{lstlisting}
Suppose that !e! is a dictionary with !n! elements.  Then the meaning
of the !sum! is the summation of !n! terms:
\begin{lstlisting}
{ @unique e2_1 -> e3_1 }  + { @unique e2_2 -> e3_2 }  + ...
\end{lstlisting}
and the !@unique!  constraints guarantees that their keys !e2_1!,
!e2_2!, !e2_3!, $\ldots$ are distinct.  Then, the outer sum will bind
the variables !k1,v1! to exactly one pair !e2_i,e3_i!.  Rule !F3!
fuses the two loops into a single loop, and uses a !let! construct to
bind !k1,v1! to !e2,e3!.

In some cases the !@unique! constraint can be inferred from the query,
but in most cases it is data dependent, and must be asserted by data
administrator when defining the TSM.  For example, the TSM for the CSR
representation of the tensor !C! in Fig.~\ref{fig:one} (c) should be
written as follows:
\begin{lstlisting}
CREATE TENSOR C AS sum(<row,_> in 0:C_len1)
    { @unique row ->
      sum(<off,col> in C_idx2(C_pos2(row):C_pos2(row+1)))
        { @unique col -> C_val(off) }
    }
\end{lstlisting}
The expression
!sum(... <(k,j), C_v> in C ...) ...!
in Fig.~\ref{fig:one} (a) desugars into two nested iterations, one for
!k! and one for !j!, and the optimizer can now use rule !F3! twice to
fuse these two iteration in TP with the two iterations in the TSM,
leading to the much more efficient program in Fig.~\ref{fig:one} (d).

%
%

\subsection{Rule Engine: Egg}

The rule engine takes as input the naive tensor plan and the
collection of rules, and repeatedly applies the rules in order to
obtain all equivalent query plans.  This task is non-trivial: the rule
engine needs to memorize all generated plans and check for duplicates,
and also needs to avoid running into an infinite loop.  Every
cost-based query optimizer that we are aware of implements its own
rule engine, which does pattern matching, duplicate detection, and
memoization of expressions.

Instead of implementing our own expression manager, we adopt a
state-of-the-art rewriting system called an Equality Saturation
(EQSAT) system~\cite{eqsat}. Specifically, we used
Egg~\cite{DBLP:journals/pacmpl/WillseyNWFTP21}.  An EQSAT system has
access to a collection of rewrite rules, and receives as input an
expression !e!.  It then constructs the plan space by maintaining a data structure, called an
e-graph, that compactly represents a set of expressions, together with
an equivalence relation over this set that can be derived from the
rules.  The e-graph consists of a set of e-classes, each e-class
consists of a set of e-nodes, and each e-node is a function symbol
with e-classes as children.

\begin{figure}
  \centering
  \includegraphics[width=\columnwidth]{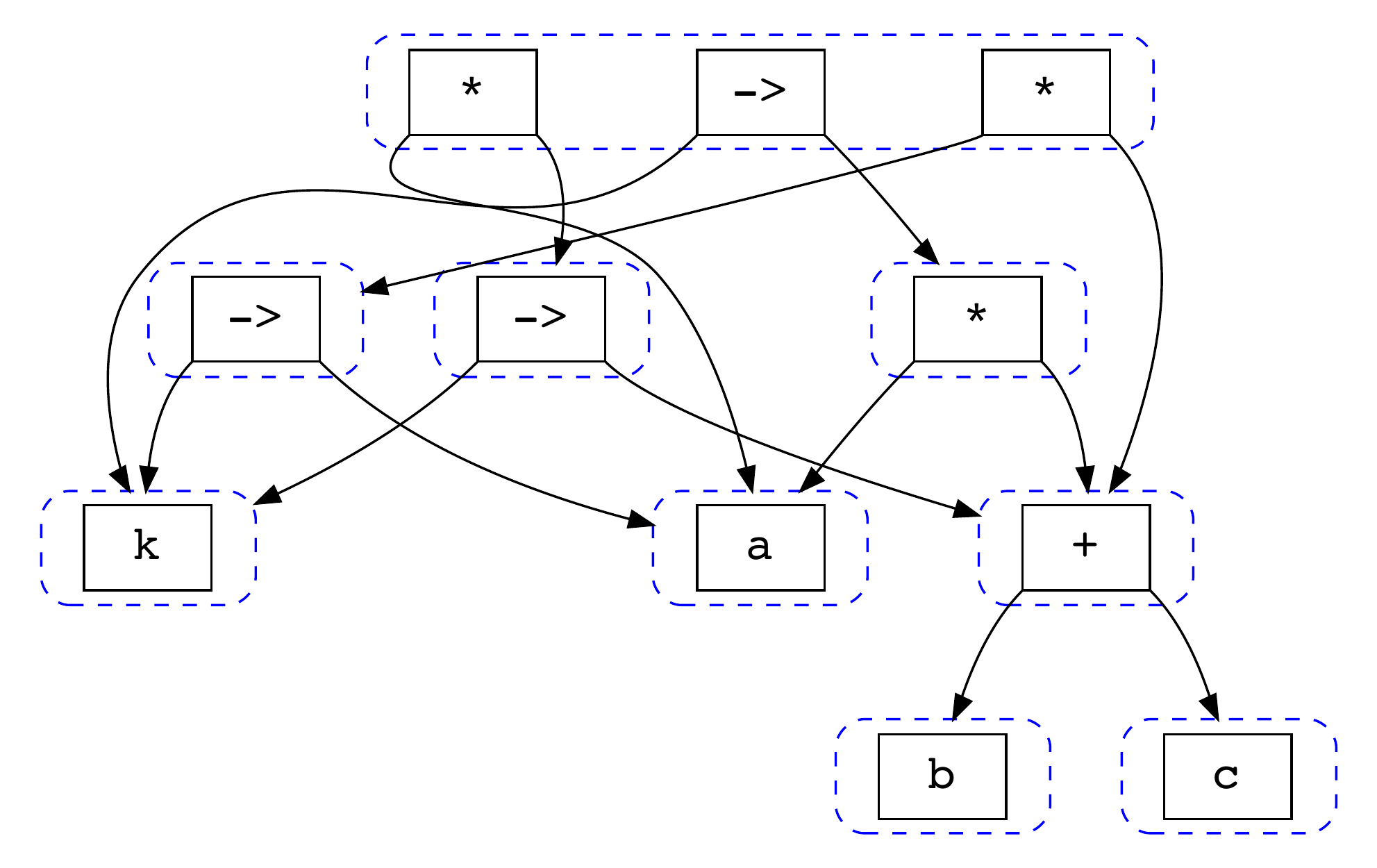}
  \caption{The e-graph of \code{a * \{ k -> b + c \}}. }
  \label{fig:egraph}
\end{figure}

For example, Fig.~\ref{fig:egraph} shows
the compact representation of all expressions equivalent to
!a * {k -> b+c}!.
The top e-class has 3 operators. The first is !->! and
has children !k! and the e-class for !a * (b+c)!, which corresponds to the associativity rule !A3!. The second is a !*! and its children
are !a! and the e-class for !{ k -> b + c }!, representing the original input.
The third is another !*!, with !{ k -> a }! and !b+c! as its children, and corresponds to the associativity rule !A2!. This e-graph corresponds to the following equivalent expressions:
\begin{lstlisting}
{ k -> a * (b+c) } = a * { k -> b+c } = { k -> a } * (b+c)
\end{lstlisting}
This e-graph is obtained by only applying the associativity rules and contains 11 nodes and 9 e-classes. By applying the rest of transformation rules (e.g., distributivity), we obtain a more complicated e-graph with 28 nodes and 15 classes.

The e-graph (i.e., plan space) is iteratively expanded by 
applying all the provided rewrite rules. This process is continued until
either the e-graph is \textit{saturated} (i.e., applying rewrite 
rules does not change the e-graph) or a threshold (e.g., number of iterations or timeout) is reached. 
Finally, Egg performs the search for the best plan through the 
\textit{extraction} procedure by a user-provided cost model.

\subsection{Managing Free Variables}

\label{subsec:vars}

A major challenge for our cost-based optimizer is that, unlike the
traditional Cascades based
framework~\cite{DBLP:journals/debu/Graefe95a}, our rules operate on a
{\em calculus} instead of an {\em algebra}.  This creates significant
challenges for managing the free variables in the expressions.

For example, consider a !let!-rule like this:
\newline
\begin{tabular}{ccc}
  !let x = e1 in e2! & \transto & !e2[e1/x]!
\end{tabular}
\newline
\noindent There are two important challenges here. First,
this rule must match with its $\alpha$-equivalent
terms, i.e., terms that become equivalent by substituting
their variable names such as !let x = e1 in x * 2! and !let y = e1 in y * 2!.
Keeping track of $\alpha$-equivalence requires a sophisticated matching
strategy that imposes scalability challenges for equality saturation.
Second, !e2[e1/x]!, which represents the result of substituting !x!
by !e1! in !e2!, is not a valid pattern in Egg.  Egg uses a compact
representation of equivalent expressions, which makes it impossible to
express substitution, since different equivalent representations of
!e2! may or may not have !x! as a free variable, or !x! may mean
different things.

We use De Bruijn indexing~\cite{de1972lambda} to provide a nameless
representation for variables. This way, the term
!let x = e1 in x * 2! is represented as
!let e1 in 
where
!
refers to the variable introduced by the closest !let!-binding.
It was shown~\cite{koehler2021sketch} that De Bruijn indexing can solve the scalability of
equality saturation by avoiding the e-graph to be overloaded with
$\alpha$-equivalent terms.

\subsection{Cardinality Estimation}

After applying all rules, \system\ uses a cardinality and cost
estimator to select the best rewriting.
We adopt ideas from~\cite{klonatos2013automatic} to represent
cardinalities of nested dictionaries.  A {\em cardinality expression}
is given by the following grammar, where 
$\cardscalar{}$ is a symbol that means that the quantity is a scalar
(e.g. has size 1), $n$ is a real number, and $\cardstatic{m}$ represents a scalar expression that stores the size $m$:
$$\card{c} ::=  \cardscalar{}  | \carddict{c}{n} | \cardstatic{m}$$
For example, if !A! is a dictionary of type
$\rangen{n_1} \rightarrow \rangen{n_2} \rightarrow \rangen{n_3}
\rightarrow \R$, then we may estimate its cardinality as
$100[10[50[\cardscalar{}]]]$, which means: for an estimated 100
indices !i!, !A(i)! is non-zero; for each such !i!, for an estimated
10 !j!'s, !A(i)(j)! is non-zero, and for each of these, for an
estimated 50 !k!'s, !A(i)(j)(k)! is non-zero.

We use the rules in Fig.~\ref{fig:cardinality} to estimate the
cardinality of a \lang\ expression.  For example, consider the
cardinality of the expressions:
\begin{lstlisting}
  sum (<i,v> in A) if (v==25) then {i -> i*3}
\end{lstlisting}
and assume that the cardinality of !A! is $1000[\cardscalar{}]$.
Further assume that the selectivity of the predicate is
!sel(v==25) = 0.02!.
Then:

\begin{tabular}{ll}
  \cardpre{}(!{i -> i*3}!) & $=1[\cardscalar{}]$ \\
  \cardpre{}(!if (v==25) then {i -> i*3}!) & $= 0.02 * 1[\cardscalar{}] = 0.02[\cardscalar{}]$ \\
  \cardpre{}(!sum(<i,v> in A) if ...!)   & $=1000 * 0.02[\cardscalar{}]=20[\cardscalar{}]$
\end{tabular}

For the cardinality of input tensors (e.g., \cardpre{}(\code{A})) and the selectivity estimates (e.g., \selpre{}(\code{e1})), 
\system{} currently relies on the information provided by DBAs or uses constants (e.g., $0.1$ for selectivity estimates). 
We leave the usage of histograms and more advanced cardinality estimation techniques for the future.

\begin{figure}
\setlength\tabcolsep{0.1cm}
\begin{tabular}{r c l}
\cardelempre{}(!e!)&=&$\card{c}$ \hfill if \cardpre{}(\code{e}) = $\carddict{\card{c}}{n}$\\
\cardsizepre{}(!e!)&=&$n$ \hfill if \cardpre{}(\code{e}) = $\carddict{\card{c}}{n}$\\ \hline
\cardpre{}(!e1(e2)!) &=& \cardelempre{}(!e1!) \\
\cardpre{}( !{ e1 -> e2 }!) &=& $\carddict{\cardpre{}(\code{e2})}{1}$ \\
\cardpre{}(\code{0:e1}) & = & $\carddict{\cardscalar}{m}$ \hfill \text{if \cardpre{}(\code{e1}) = $\cardstatic{m}$}\\
\cardpre{}(!let x = e1 in e2!) &=& \cardpre{}(!e2!)\\
\cardpre{}(!if(e1) then e2!) &=& $\begin{cases}
               \cardscalar&\text{if \cardpre{}(\code{e2}) = $\cardscalar$}\\
               \carddict{\card{c}}{\text{\selpre{}(\code{e1})}\cdot n}&\text{if \cardpre{}(\code{e2}) = $\carddict{\card{c}}{n}$}
            \end{cases}$\\
\cardpre{}(!sum(<k,v> in e1) e2!) &=& $\begin{cases}
               \cardscalar&\text{if \cardpre{}(\code{e2}) = $\cardscalar$}\\
               \carddict{\card{c}}{\cardsizepre{}(\code{e1})\cdot n}&\text{if \cardpre{}(\code{e2}) = $\carddict{\card{c}}{n}$}
            \end{cases}$\\
\end{tabular}

\caption{Cardinality estimation rules.}
\label{fig:cardinality}
\end{figure}

\subsection{Physical Plans}
\label{sec:physical}

So far all expressions in \lang\ are logical plans.  We describe here
how we convert \lang\ expressions into physical plans, which we
execute on our runtime system, Julia.  Simple scalar operators like
!a+b! or !a*b! get converted immediately into physical operations.
Julia also supports plus and times operators on dictionaries
(tensors); if that were not the case, then we can force the optimizer
to write such operations explicitly as loops, e.g. we rewrite the
expression !a*b!, where !a! is a scalar and !b! is a dictionary, into:
\begin{lstlisting}
sum (<i,vb> in b) { i -> a*vb }
\end{lstlisting}
and assign a cost of $\infty$ to !+! and !*! operators applied to
dictionaries.

The physical operator associated to
!sum (<k,v> in e1) e2!
is a !for! loop iterating over the dictionary !e1!.  To make this loop
concrete, \system\ needs to know how the dictionary !e1! is
represented.  In our system, there are two choices: as a dense vector,
or as a hash-map.  \system\ knows the type of storage for the arrays,
hash maps, and tries of the physical storage, since they were
explicitly declared in the TSM.  For all other constructed
dictionaries, \system\ needs to choose whether to construct a dense
vector, or a hash map.  We do this by adding the following two rules
to the collection of rules:

\begin{tabular}{lll}
  !{k -> e}! & \transto & !{ @dense k -> e}! \\
  !{k -> e}! & \transto & !{ @hash k -> e}!
\end{tabular}

In the first rule !{k -> e}!  becomes an entry of a dense array,
in the second rule it becomes an entry of a hash-map.  We assign a
cost of $\infty$ to any expression that still contains a logical
dictionary !{k -> e}!, thus forcing the optimizer to choose either a
dense array or a hash-map representation.

Finally, we add one additional physical operator to \lang:

\begin{lstlisting}
     merge(<k1,k2,v> in <e1,e2>) e3
\end{lstlisting}

Both !e1! and !e2! must be dictionaries of real values, in other words
they must be vectors, and, in that case, the semantics of !merge! is:
\begin{lstlisting}
  sum(<k1,v> in e1, <k2,u> in e2) (if (v==u) then e3)
\end{lstlisting}

This is captured by the Fusion Rule !F4! in
Fig.~\ref{fig:opt_rules}.

\subsection{Cost Estimate}
\label{sec:cost}

\begin{figure}
\setlength\tabcolsep{0.02cm}
\begin{tabular}{r c l}
\costpre{}(!e1(e2)!) &=& \costpre{}(!e1!) $ + $ \costpre{}(!e2!) $ + \coef{lookup}$(!e1!)\\
\costpre{}(!{ e1 -> e2 }!) &=& $\infty$\\
\costpre{}(!{@dense e1 -> e2 }!) &=& \costpre{}(!e1!) $ + $ \costpre{}(!e2!) $ + \coef{arr-insert}$(!e1!, !e2!)\\
\costpre{}(!{@hash e1 -> e2 }!) &=& \costpre{}(!e1!) $ + $ \costpre{}(!e2!) $ + \coef{hash-insert}$(!e1!, !e2!)\\
\costpre{}(!let x = e1 in e2!) &=& $\coef{mater}$(!e1!)$\cdot$\costpre{}(!e1!) $ + $ \costpre{}(!e2!)\\
\costpre{}(!if(e1) then e2!) &=& \costpre{}(!e1!)$ + $\selpre{}(!e1!)$\cdot$\costpre{}(!e2!)\\
\costpre{}(!sum(<k,v> in e1) e2!) &=&
\costpre{}(!e1!)$ + \coef{iter}$(!e1!)$\cdot \cardsizepre{}(\code{e1})\cdot$\costpre{}(!e2!)\\
\end{tabular}

\begin{tabular}{l}
\costpre{}(!merge(<k1,k2,v> in <e1,e2>) e3!) = \\
\costpre{}(!e1!)$ + $\costpre{}(!e2!)$ + (\coef{iter}$(!e1!)$\cdot \cardsizepre{}(\code{e1})+\coef{iter}$(!e2!)$\cdot \cardsizepre{}(\code{e2}))\cdot$\costpre{}(!e3!)
\end{tabular}
\vspace*{-1em}
\caption{Cost estimation rules.}
\label{fig:cost}
\vspace*{-1.5em}
\end{figure}

Finally, the cost of a physical plan is estimated using the rules
shown in Fig.\ref{fig:cost}. These inference rules include parameters
that are dependent on the type of the underlying collection (e.g.,
$\coef{lookup}$ and $\coef{iter}$ for a dense-array is smaller than the one for a hash-map).
We notice that a logical plan for which
we have not chosen between a dense array and hash map will have cost
$\infty$.



\begin{table}
  \begin{tabular}{l l r r}
    \hline
    Tensor    & Dimensions                      & Density             & \# non-zeros \\ \hline \hline
    cant      & 62K $\times$ 62K                & $1 \times 10^{-3}$  & 2.03M        \\
    consph    & 83K $\times$ 83K                & $9 \times 10^{-4}$  & 3.05M        \\
    cop20k\_A & 121K $\times$ 121K              & $2 \times 10^{-4}$  & 1.36M        \\
    pdb1HYS   & 36K $\times$ 36K                & $3 \times 10^{-3}$  & 2.19M        \\
    rma10     & 46K $\times$ 46K                & $1 \times 10^{-3}$  & 2.37M        \\
    webbase   & 1M $\times$ 1M                  & $3 \times 10^{-6}$  & 3.11M        \\
    \hline
    NIPS      & 2.4K $\times$ 2.8K $\times$ 14K & $3 \times 10^{-5}$  & 31.31M       \\
    NELL      & 12K $\times$ 9.2K $\times$ 29K  & $2 \times 10^{-5}$  & 76.88M       \\
    Facebook  & 1.6K $\times$ 64K $\times$ 64K  & $1 \times 10^{-7}$  & 0.74 M       \\
    Enron     & 6K $\times$ 5.7K $\times$ 244K  & $3 \times 10^{-6}$  & 3.10 M       \\
    \hline
  \end{tabular}
  \caption{Real-world matrices and rank-3 tensors used in the experiments.}
  \label{tbl:datasets}
  \vspace*{-3em}
\end{table}

\begin{table*}
\setlength\tabcolsep{1pt}
\begin{footnotesize}
  \begin{tabular}[]{| l | c | c | c | c | c | c | c | c | }\hline
    Tensor Program & \system{} / \taco{}       & \scipy{}   & \numpy{} & \pytorch{} & \tensorflow{}   & \duckdb{} & Sec.~\ref{sec:exp:e2e} Dim. & Sec.~\ref{sec:exp:sf} Dim.    \\ \hline \hline
    \expmmm{}: $Q(i, j) = \sum_{k} A(i, k) \cdot B(k, j)$                       & CSR, CSR      & CSR, CSR   & Dense, Dense & CSR, Dense & COO, Dense & COO, COO & $B$: $\_\times250$ & $A$:$10^3\times10^3$,$B$:$\_\times10^3$      \\\hline
    \expsmmm{}: $Q() = \sum_{i,j,k} A(i, k) \cdot B(k, j)$                       & CSC, CSR      & CSR, CSR   & Dense, Dense & CSR, Dense & COO, Dense & COO, COO & $B$: $\_\times250$ & $A$:$10^5\times10^5$,$B$:$\_\times10^5$    \\\hline
    \expbatax{}: $Q(j) = \sum_{i,k} \beta \cdot A(i, j) \cdot A(i, k) \cdot X(k)$ & CSR, Dense   & CSR, Dense & Dense, Dense & CSR, Dense & COO, Dense & COO, COO & $X$: $\_$ & $A$:$10^5\times10^5$,$X$:$\_$    \\\hline
    \expttm{}: $Q(i, j, k) = \sum_{l} A(i, j, l) \cdot B(k, l)$                 & CSF, CSC/CSR     & ---        & ---  & --- & ---         & COO, COO & $B$: $\_\times25$ & --- \\\hline
    \expmttkrp{}: $Q(i, j) = \sum_{k,l} A(i, k, l) \cdot B(k, j) \cdot C(l, j)$    & CSF, CSR, CSC & ---        & --- & --- & ---          & COO, COO, COO & $B$: $\_\times25$ & --- \\\hline
  \end{tabular}
\end{footnotesize}
  \caption{Tensor Programs and their best storage formats for
   each system. \scipy{}, \numpy{}, \pytorch{}, and \tensorflow{} do not support higher-order sparse tensors. \duckdb{}
   encodes the tensors as relations, which are comparable to the coordinate (COO) format in
   tensor systems. The missing dimensions, denoted by $\_$ or not included, can be inferred from the context (e.g., for Sec.~\ref{sec:exp:e2e} the dimension of $A$ is specified in Table~\ref{tbl:datasets} and the number of rows of $B$ is the same as the number of columns of $A$).}
  \label{tbl:tps}
  \vspace*{-1em}
\end{table*}

\section{Experiments}
\label{sec:experiments}

In this section we present an empirical evaluation of \system, by
running on several common tensor kernels, with a variety of real and
synthetic matrices and tensors, and comparing it with six other
systems. We studied the following questions:
\begin{enumerate}
  \item How much do tensor programs over flexible storage benefit from cost-based optimization?
  \item How do different choices of storage formats for different data
        sparsities affect the run-time performance, and does \system\ take
        best advantage of the given storage format?
  \item How much do specific sets of rewrite rules contribute to the
        optimization?  In particular we would like to understand the
        contribution of loop fusion and factorization.
  \item How complex is the optimization task?  How many applications of
        rules are needed to optimize Tensor Programs?
  \item How practical is the optimization process?
  Does the run time improvement outweight the optimization overhead?
\end{enumerate}

In addition, we discuss our experience with using Egg as our rule rewrite system at the end
of the section.

\begin{figure*}[t]
  \includegraphics[width=0.32\textwidth]{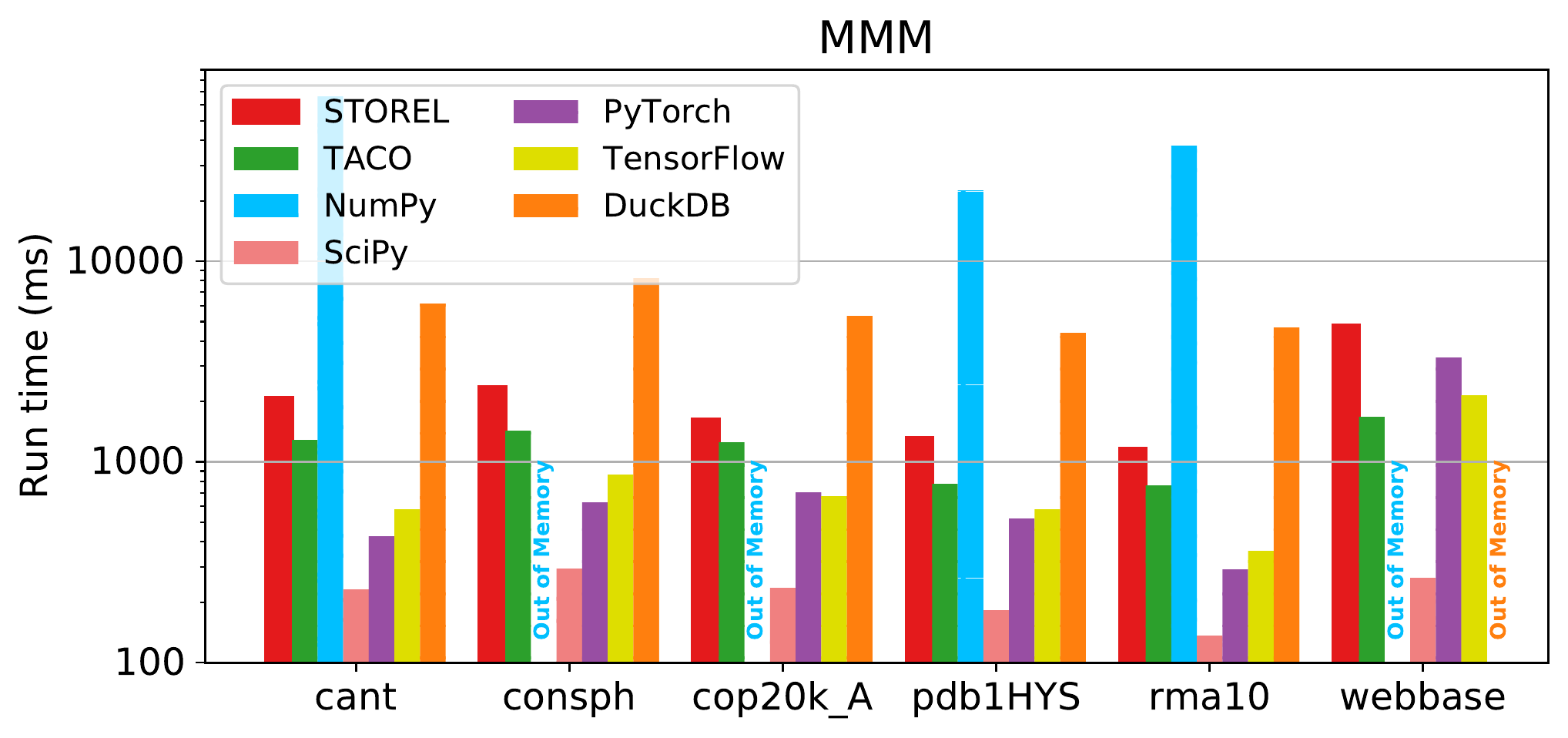}~%
  \includegraphics[width=0.32\textwidth]{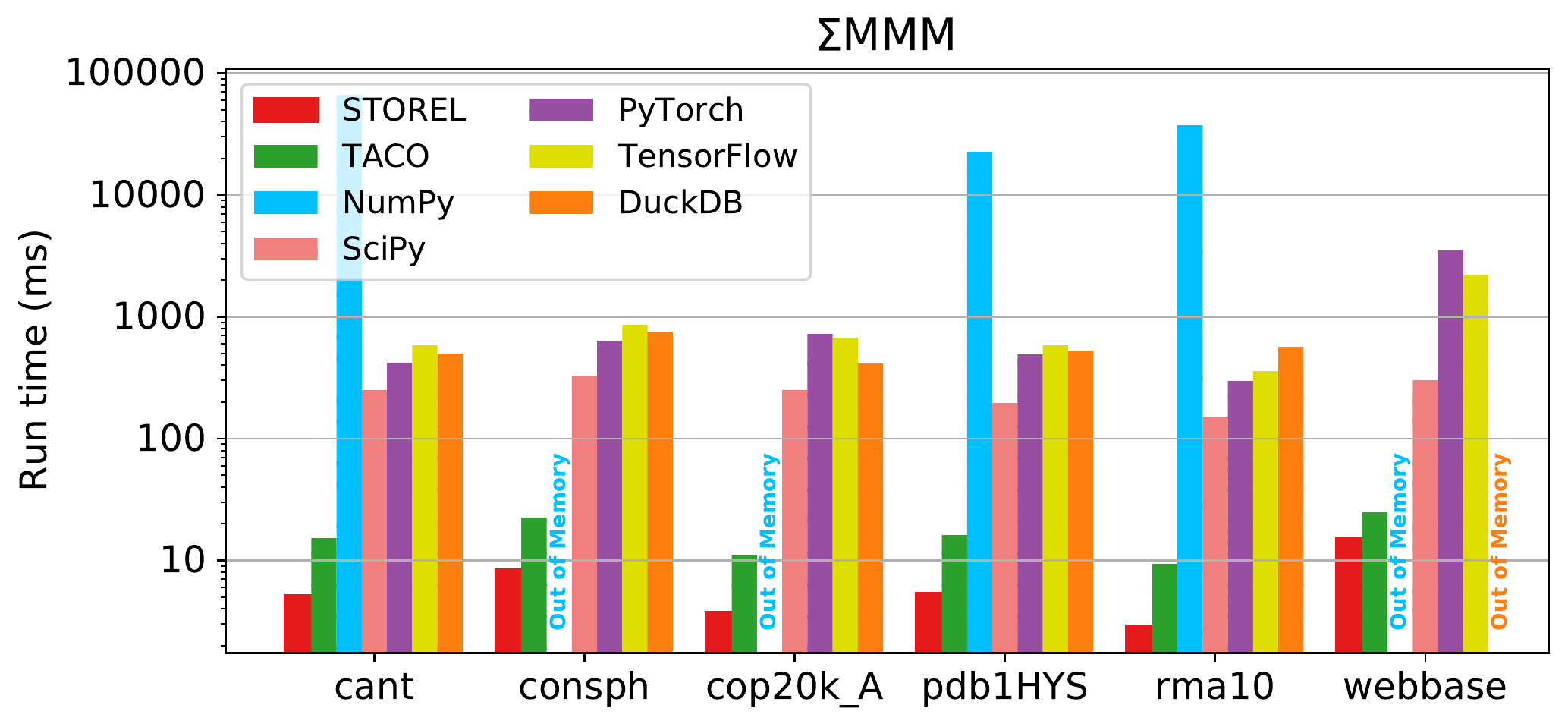}~
  \includegraphics[width=0.32\textwidth]{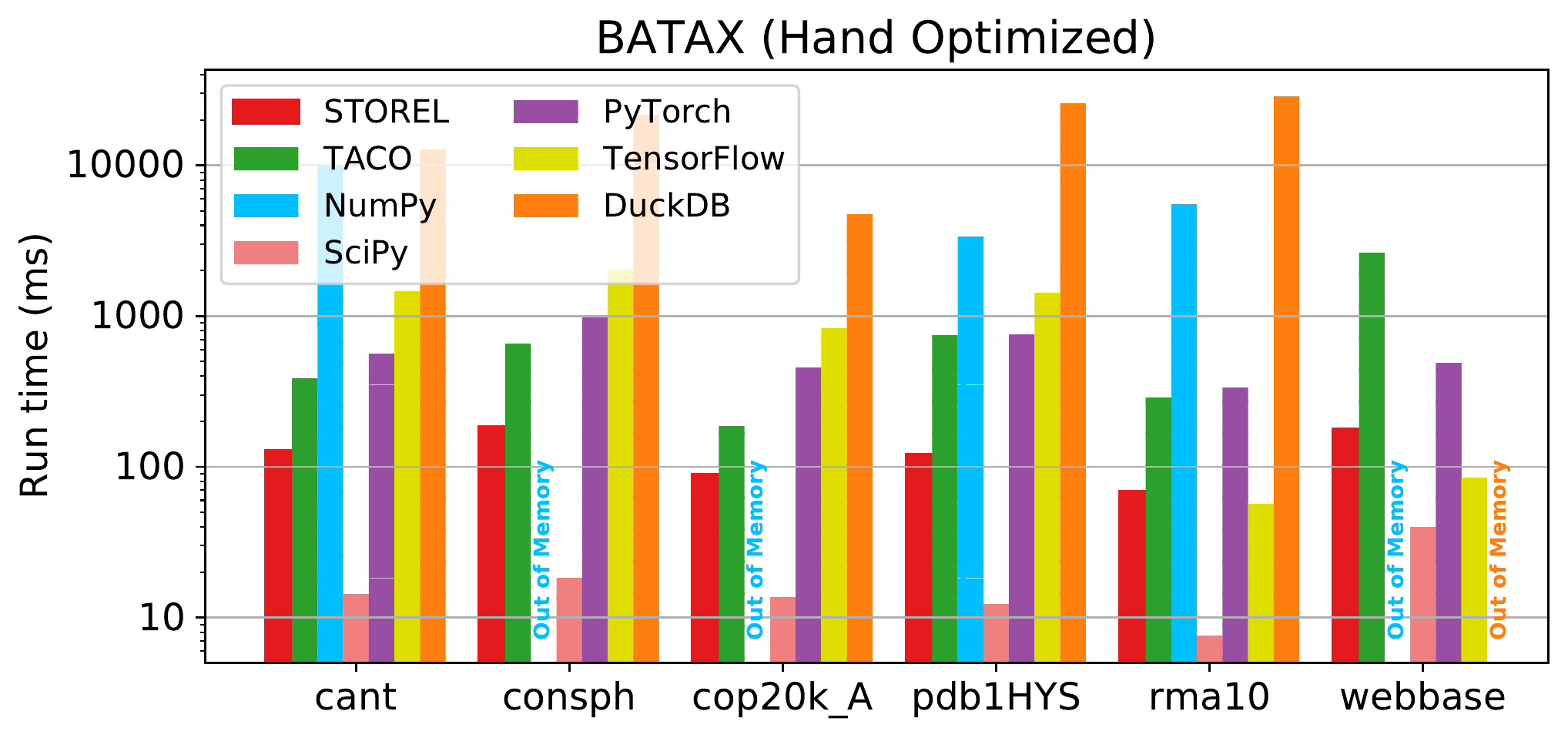}

  \includegraphics[width=0.32\textwidth]{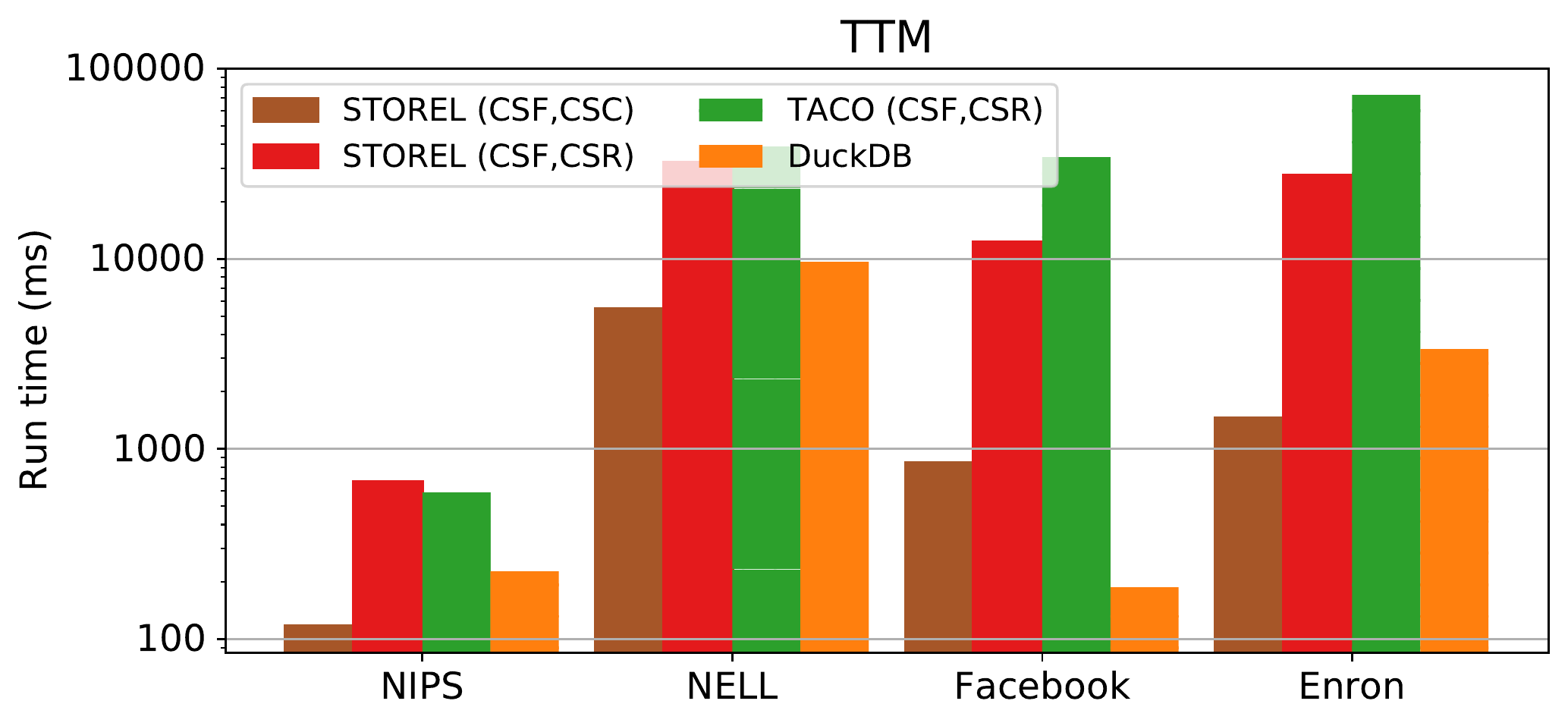}~
  \includegraphics[width=0.32\textwidth]{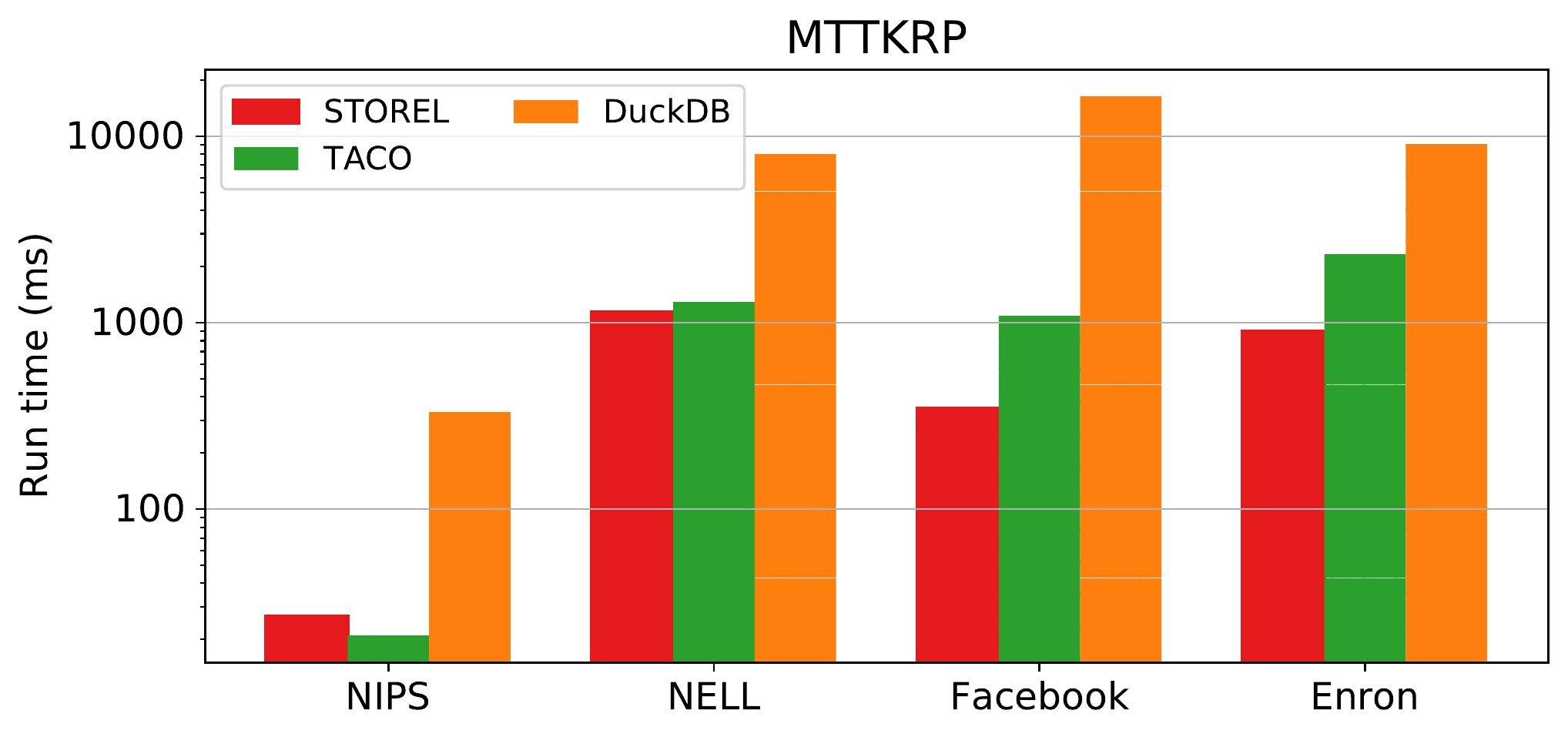}~
  \includegraphics[width=0.32\textwidth]{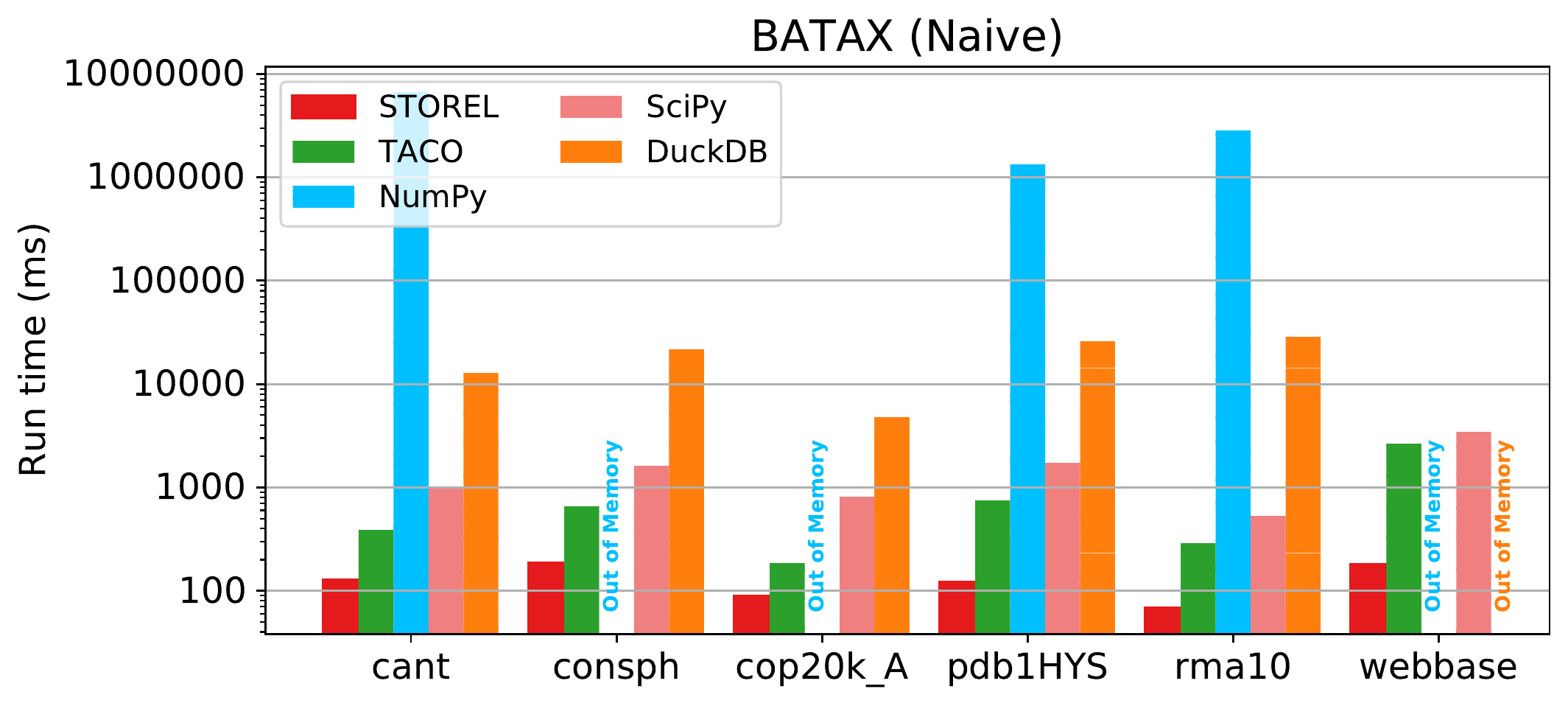}
  \vspace{-0.4cm}
  \caption{ End-to-end run time (in milliseconds) for \system, \taco, \scipy, \numpy, PyTorch, TensorFlow, and
    \duckdb{} for different kernels and real-world matrices and tensors. }
  \label{fig:exp:e2e}
  \vspace{-0.4cm}
\end{figure*}

\smartpara{Experimental Setup}
We conducted our experiments on an AWS t2.2xlarge instance with 8 vCPUs, 32 GBs of RAM, and Ubuntu 22.04 LTS. Our system uses Julia 1.7.3~\cite{julia} for
executing the generated code. The other systems we benchmark are \taco{}~\cite{taco},
\numpy{} 1.22.3~\cite{numpy}, \scipy{} 1.8.1~\cite{scipy}, \pytorch{} 1.11.0, \tensorflow{} 2.9.1, and \duckdb{}
0.3.2~\cite{DBLP:conf/sigmod/RaasveldtM19}. We use G++ 11.2.0 to compile the
generated C code in  \taco{}, and use Python 3.10.4 for \numpy{}, \scipy{}, \pytorch{}, and \tensorflow{}. In \duckdb{},
all tensors are encoded as relations, which are comparable to the coordinate (COO) format in
tensor systems, and  we provide all relevant indices.  In addition, we use an in-memory
database and the Python API to interact with \duckdb{}. Python-based frameworks do not support
sparse tensors with more than two dimensions, so we only report the times for kernels that
only contain matrices and vectors. In addition, \numpy{} only supports dense storage
formats.

All experiments are run on one CPU core, and we report the average
execution time of five runs.  In all cases we measure only the
execution time (which includes the assembly time for \taco), and we
exclude from the run time the creation and indexing of storage format,
loading time, compilation time, and optimization times respectively,
for the systems that have these components.

\smartpara{Datasets}
We use both real world and synthetic datasets. For the former, we collected six sparse
matrices from the SuiteSparse Matrix Collection~\cite{davis2011university}, and four rank-3
tensors from the FROSTT Tensor Collection~\cite{smith2017frostt}. Table~\ref{tbl:datasets}
presents a summary of these datasets. For synthetic data, we generate random matrices and
vectors with specified sparsity and dimensions.

\smartpara{Workloads}
Table~\ref{tbl:tps} presents the tensor programs we consider in this evaluation. \expmmm{}
stands for matrix-matrix multiplication; \expsmmm{} computes the summation over a
matrix-matrix multiplication; \expbatax{} was previously studied in~\cite{bto}; \expttm{}
computes the tensor times matrix multiplication; and \expmttkrp{} stands for matricized
tensor times Khatri-Rao product. Both \expttm{} and \expmttkrp{} have been studied
extensively in papers on \taco{} (see e.g., \cite{taco,taco-formats}).

\subsection{Benchmarking Tensor Programs}
\label{sec:exp:e2e}

Here we addressed the first question: how much do tensor programs over flexible storage
benefit from the application of rewrite rules?  We present the benchmark of \system{},
\taco{}, \scipy{}, \numpy{}, \pytorch{}, \tensorflow{}, and \duckdb{} on all considered tensor programs.

\smartpara{Storage Formats} Table~\ref{tbl:tps} presents the best storage formats we found for each considered tensor program and system. For each experiment, the $A$ matrix or tensor in the
respective kernel is defined by one of the datasets in Table~\ref{tbl:datasets}. All other
matrices are synthetically generated with sparsity $2^{-5}$.

\duckdb{} and \numpy{} only support a single storage format, i.e., relations and
respectively dense matrices/vectors.  For that reason, we only
consider them for the kernels that operate on matrices and vectors. For \scipy{} and \pytorch{} we use the CSR format for all matrices,
because our experiments with CSC matrices were consistently slower. For \expttm, we report
the performance for two storage formats in \system{}. The first uses a CSF tensor and a CSC
matrix, which is the optimal storage specification for this kernel. \taco{}, however, fails
to compile the kernel with the CSC matrix, which we reported to the \taco{} developers.
Thus, we also report the performance for a CSF tensor and a CSR matrix for a direct
comparison of \taco{} and \system{}.
Finally, \pytorch{} and \tensorflow{} have a limited support
for sparse matrix operations\footnote{\pytorch{} and \tensorflow{} only support a sparse-dense matrix multiplication.}. Thus, we include the results for a hand-optimized plan for the \expbatax{} kernel.

\smartpara{Results} Figure~\ref{fig:exp:e2e} presents our run-time benchmarks for the above
workloads.
\system{} is always at least competitive with \taco{},
and achieves significant performance improvements for kernels that
benefit from our factorization rewrite rules. This is the case for the
\expsmmm{}, \expbatax{}, and \expmttkrp{} kernels. For instance,
\system{} can compute \expbatax{} up to $16.4\times$ faster than
\taco{} for webbase. Thus, our rewrite rules can lead to significant
performance improvements for a variety of tensor programs.

The \expmmm{} benchmark is a simple matrix multiplication and offers
almost no opportunity for optimization, but instead is a good
benchmark for comparing the physical runtimes of the systems.
\scipy{} has the best run time of all, while those of \system\ and
\taco\ are comparable.
\scipy{} has a suite of highly optimized
low-level primitives, like sparse-sparse matrix multiplication.
\pytorch{} and \tensorflow{}, however, support sparse-dense
matrix multiplication, and thus show a worse performance for \expmmm{}.
These frameworks require the
composition of such primitives with costly materialization of
intermediate results for the other benchmarks.
We observe that
\system\ can be up to two orders of magnitude faster than them when
high-level optimizations are possible.
However, for hand-optimized plans (e.g., \expbatax{}) the highly optimized primitives show better performance than the Julia-based runtime of \system{}.
\numpy{} requires all inputs to be dense, and runs out of memory for all but four
experiments, where  \system{} outperforms it by two orders of
magnitude.  This  exemplifies the importance of flexible storage.

\duckdb{} uses quite different physical operators, and a direct comparison of the wall clock
time is not very informative. We observe, however, that \duckdb{} is remarkably efficient
for the kernels that do not offer opportunities for cost-based optimization. For
instance, \duckdb{} has excellent performance for the \expttm{} kernel, which translates
into a simple aggregate-join query. In contrast, \duckdb{} is significantly slower for
the  \expsmmm{}, \expbatax{} and \expmttkrp{} kernels. For  \expsmmm{}, this is because
\duckdb{} does not push the summation past the join. For \expbatax{} and \expmttkrp{}
kernels, \duckdb{} is not able to factorize the computation, and uses binary join plans
which construct costly intermediate results.

\begin{figure*}[t]
  \includegraphics[width=0.32\textwidth]{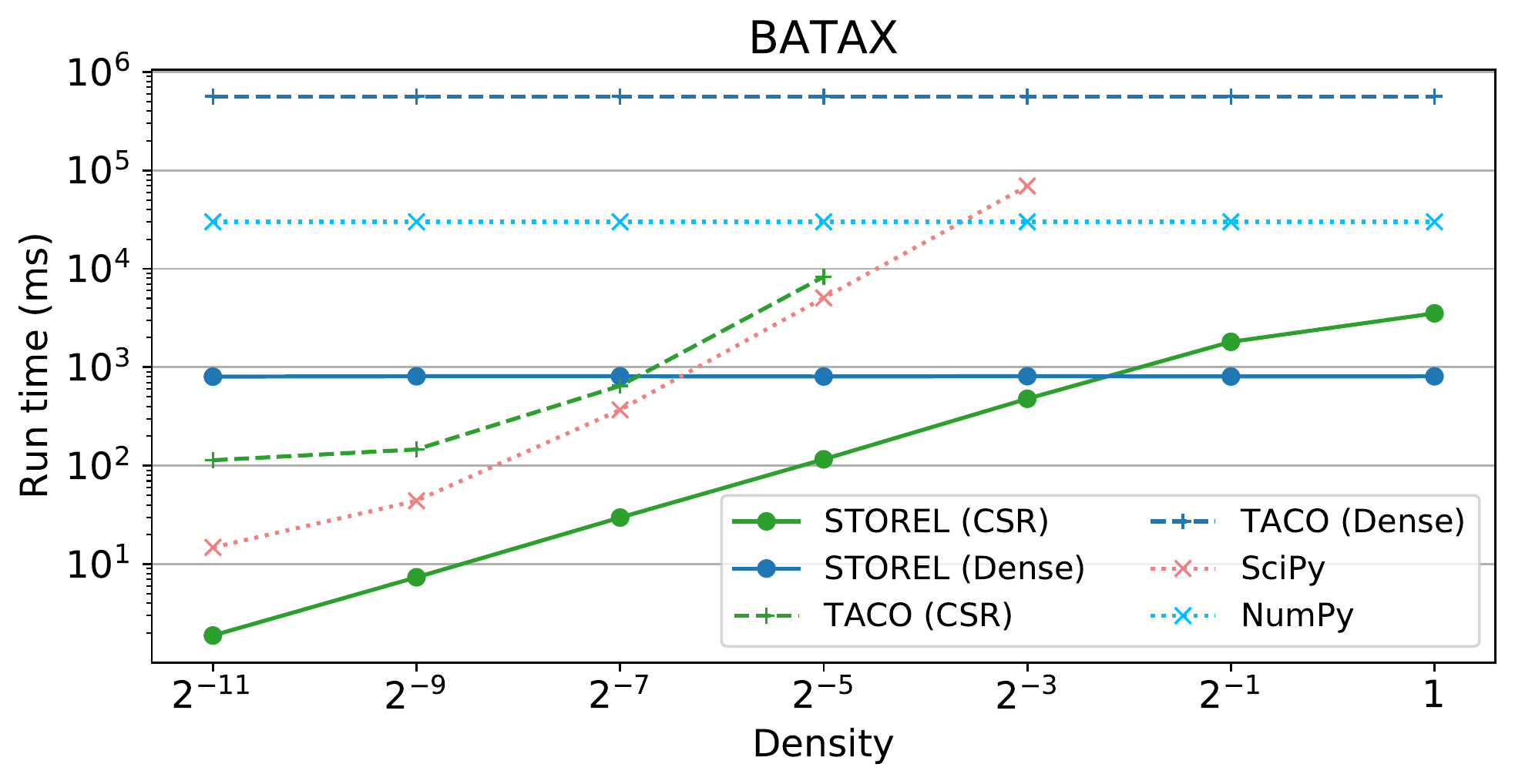}~
  \includegraphics[width=0.32\textwidth]{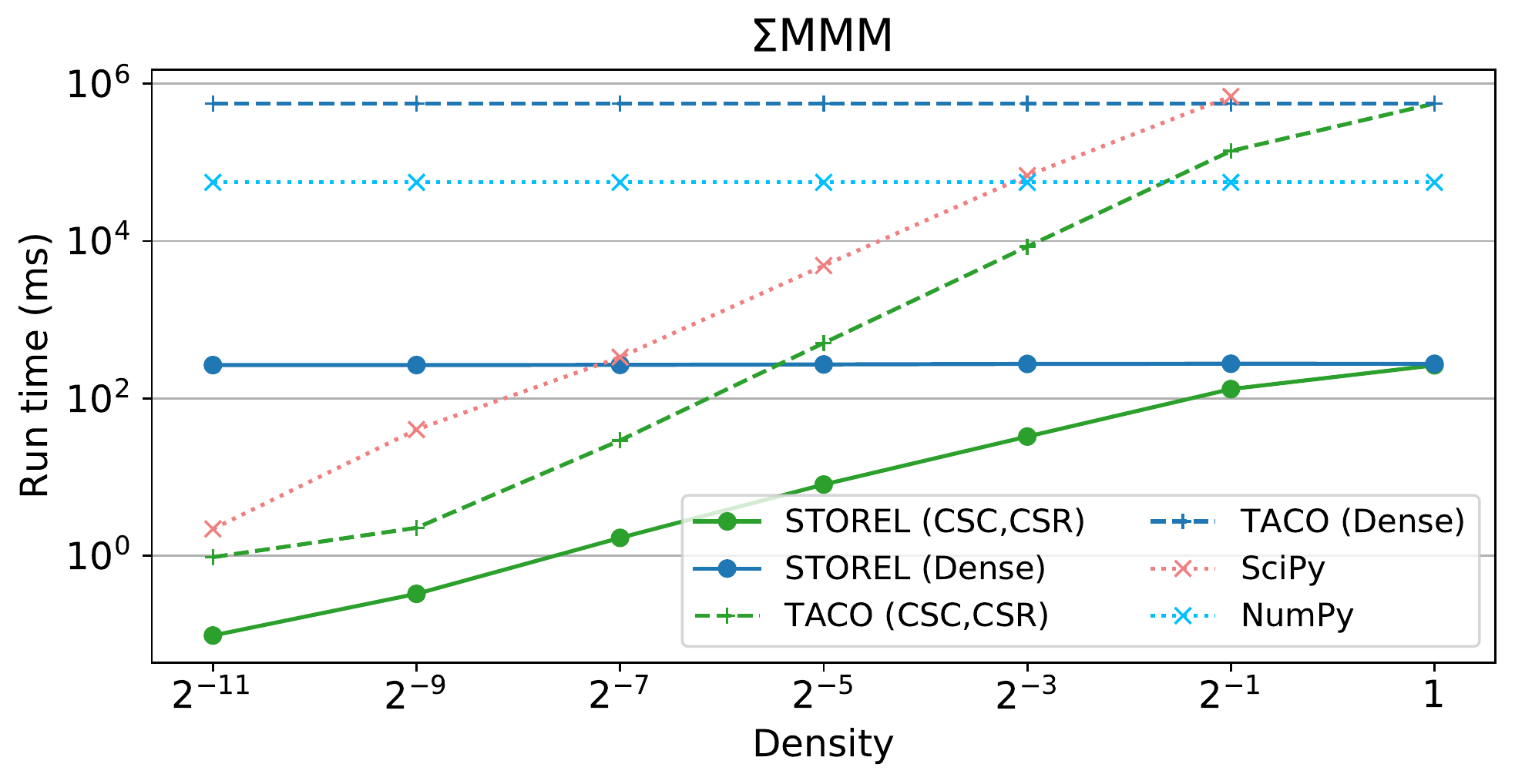}~
  \includegraphics[width=0.32\textwidth]{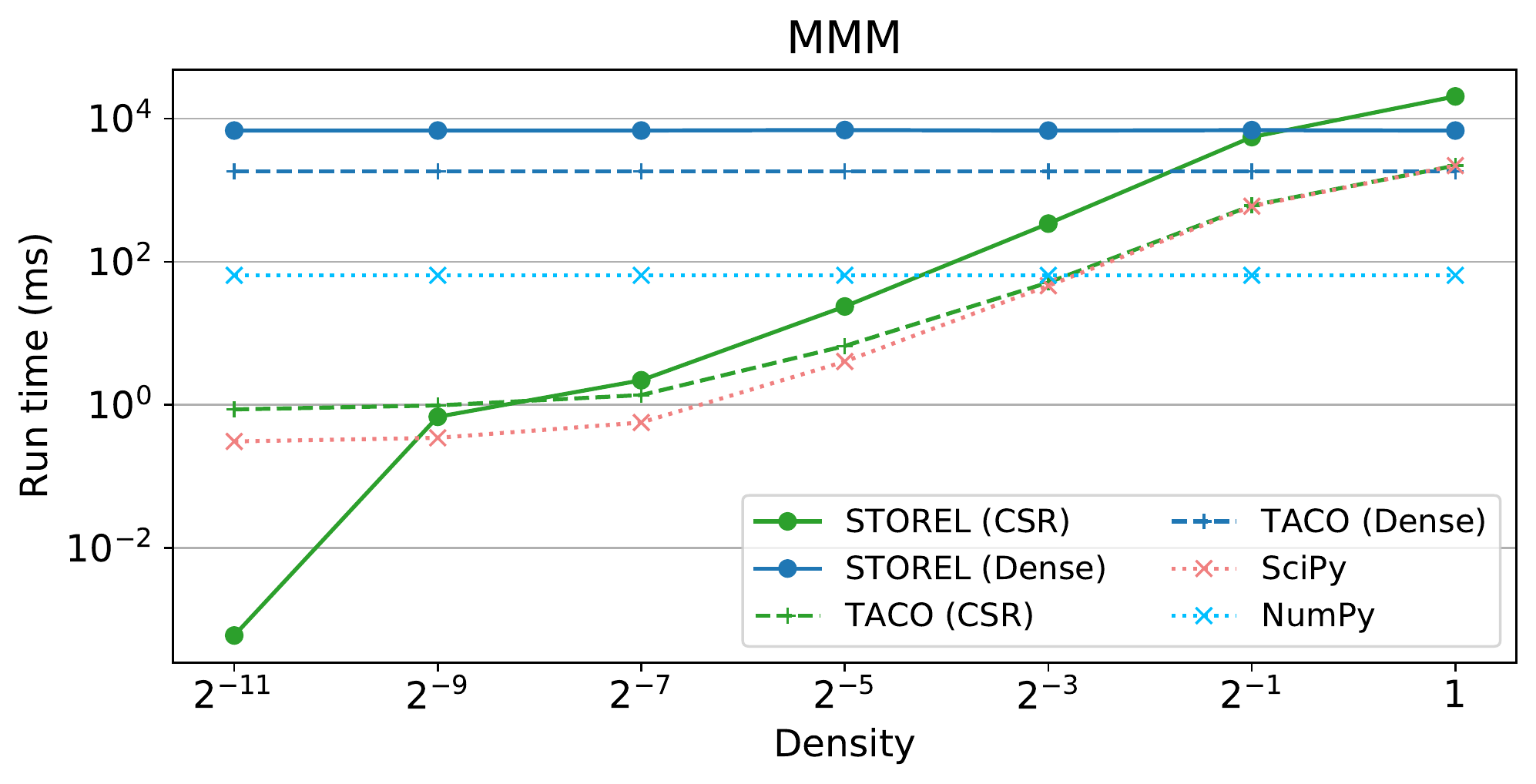}~
  \vspace{-0.4cm}
  \caption{Runtime of \system{}, \taco{}, \scipy{}, and \numpy{} for varying sparsity using
    sparse and dense storage formats.}
  \label{fig:exp:sparsity}
  \vspace{-0.4cm}
\end{figure*}

\subsection{Effect of the Storage Mapping}
\label{sec:exp:sf}

Next, we turn our attention to the second question: do different choices of storage format
for different data sparsities affect the run-time performance, and does \system\ take best
advantage of the given storage format?

We consider the \expbatax{}, \expsmmm{}, and \expmmm{} kernels and present the run time for \system{},
\taco{}, \numpy{}, and \scipy{} for different sparsity factors in the input matrices. For
\system{} and \taco{}, we further consider both the sparse storage format as in
Sec.~\ref{sec:exp:e2e}, as well as the fully dense storage format. We only use synthetic
datasets for this benchmark. For \expsmmm{} and \expmmm{}, we vary the sparsity in both matrices, and we
use the same sparsity factor for both. For \expbatax{}, we consider the naive plan and only the matrix varies the sparsity,
while the vector remains dense.

The results are presented in Figure~\ref{fig:exp:sparsity}. We observe that \system{} adapts
to the given storage format: the sparse variant is more efficient in most cases, at high
densities the dense format becomes more efficient, as expected.  Note that for \expsmmm{} and \expbatax{} \system{}
outperforms all other systems independent of the sparsity, due to the factorization rules.
However, for \expmmm{} the low-level primitives of \numpy{} and \scipy{} outperform the nested loops generated by \system{} and \taco{}.
As an example, for higher densities \numpy{} outperforms all competitors, thanks to the heavily-tuned low-level primitive provided by BLAS.
We leave the synthesis of such primitives (e.g., BLAS routines), instead of the nested loops, for future.

\subsection{Effect of Rewrite Rules}

\label{subsec:impact}

\begin{figure}[t]
  \includegraphics[width=\columnwidth]{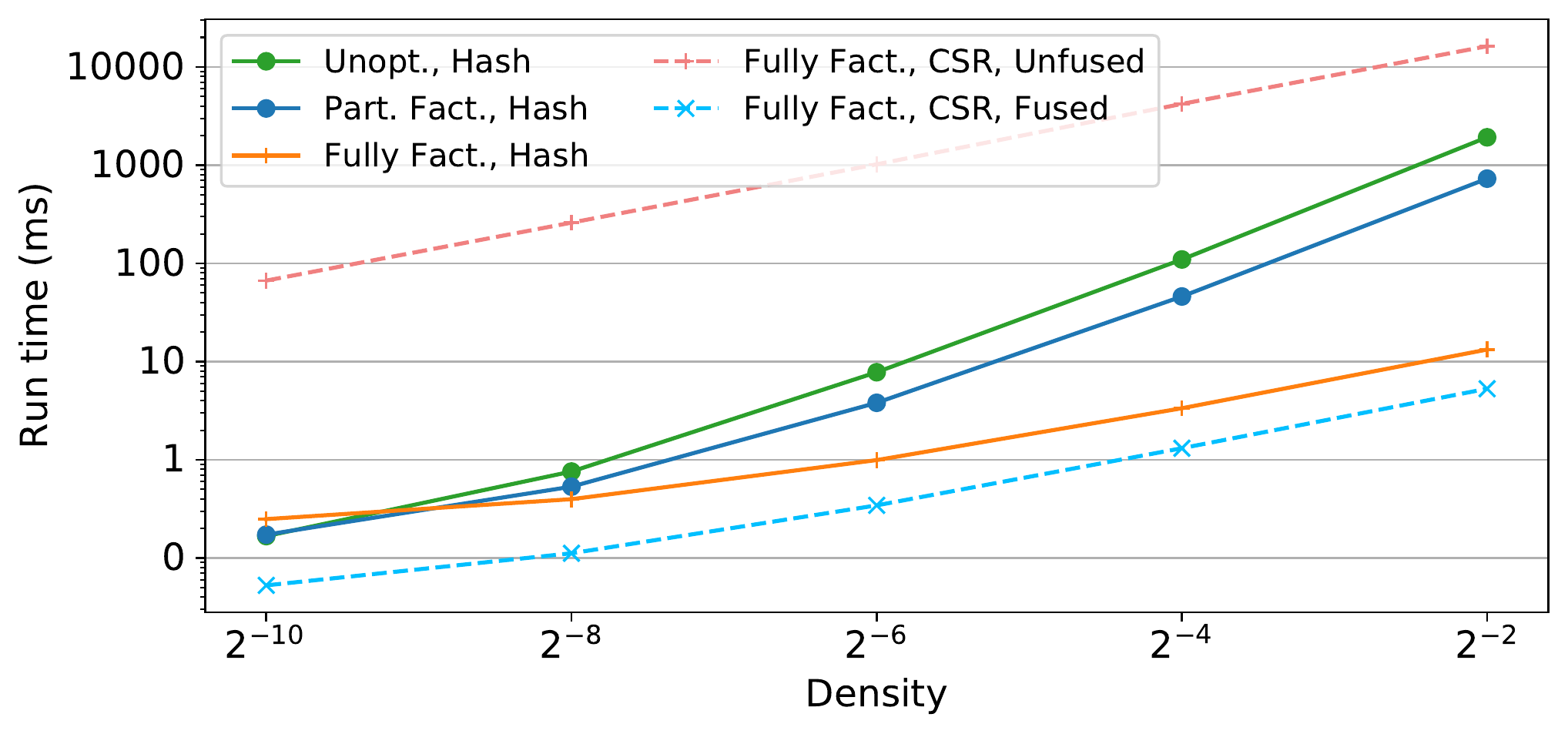}
  \vspace{-0.9cm}
  \caption{Impact of factorization and fusion rules on the \expbatax{} kernel. The dimension of matrix $A$ is $10^3\times10^3$. }
  \label{fig:exp:opts}
  \vspace{-0.4cm}
\end{figure}

Here we address the  third question: study the contribution of two
classes of rewrite rules, loop fusion and factorization, on the
overall optimization.
For that purpose we use the \expbatax{} kernel as an example. The
results are presented in Figure~\ref{fig:exp:opts}.

We first consider the case where the input matrix is a nested hash-map (trie), in which case we only benefit from
the factorization rules. The following expression presents the unoptimized program, which we
use as the baseline (the green line in Figure~\ref{fig:exp:opts}):
\begin{lstlisting}
  sum(<i, Ai> in A)
    sum(<j, Aij> in Ai)
      sum(<k, Aik> in Ai)
        { j -> beta * Aij * Aik * x(k) }
\end{lstlisting}

This kernel has two factorization opportunities. The first rewriting hoists the construction
of the dictionary with key $j$ out of the inner sum:
\begin{lstlisting}
sum(<i, Ai> in A)
  sum(<j, Aij> in Ai)
    { j -> sum(<k, Aik> in Ai)
      beta * Aij * Aik * x(k) }
\end{lstlisting}
The rewritten kernel, represented by the blue line, is between one to two orders of
magnitude faster than the non-optimized kernel, depending on the sparsity.

The second factorization opportunity hoists the inner sum over $k$ outside the sum over $j$:
\begin{lstlisting}
  sum(<i, Ai> in A)
    let t = (sum(<k, Aik> in Ai) Aik * x(k))
    in (sum(<j, Aij> in Ai) { j ->  beta * Aij * t})
\end{lstlisting}
This optimization can further improve the run time by an order of magnitude. For very sparse
data, however, it is more beneficial to avoid hoisting the loop outside. This is because the
inner sum may not be executed at all for many $i$ values.

We further consider the case where the matrix is stored with a CSR format, in which case there is a
fusion opportunity. The two dashed lines in Figure~\ref{fig:exp:opts} represent the run time
of \system{} with and without the fusion of the CSR matrix, while at the same time
exploiting both factorization opportunities as described above. We observe that the unfused
variant comes with heavy overhead, because the program first materialized the matrix that is
defined by the storage representation and then executes the program. This would be $2\times$
worse than the non-optimized baseline, despite the use of factorization. It is only with the
fusion of the storage representation and the actual program that we achieve the best
performance, which is $3\times$ faster than the optimized hash-based implementation.

\subsection{Cost and Complexity of the Rewrite-based Optimization}

Finally, we address here the fourth question: what is the cost and the
complexity of the cost-based optimization?  Recall that we have not
included the optimization cost in our experiments so far.

Our rewrite rules define a huge search space, and it proved to be too large for the current version of Egg to
saturate. Our solution was to restrict the search space by
splitting our optimization pipeline into two stages. First, we apply our rewrite rules to the
tensor program without taking the storage format into account. Then, we further optimize the
resulting program in conjunction with the provided storage format.
We notice that most Cascade-style optimizers also partition  the
optimization into several stages, in order to reduce the search space
and make the optimization possible.

Table~\ref{tbl:egg:metrics} presents the key metrics for the two optimization passes in Egg
per tensor program. We observe that, even with the separation of the optimization pipeline,
Egg explores a large search space and constructs an e-graph with tens of thousands of
equality classes. As a result, the optimization time can take up to 1.7 seconds in total,
which is longer than the execution time of the kernel for small tensors. In the next section, we
investigate in more detail the trade-off between optimization and run time.

\begin{table}
  \begin{small}
  \begin{center}
    \begin{tabular}{|l|r|r|r|r|r|r|c|}
      \hline
      \textbf{Kernel}             & \textbf{Time (ms)} & \textbf{Iters.} & \textbf{Nodes} & \textbf{Classes} & \textbf{Memos} \\ \hline
\multirow{2}{*}{\expbatax} & 445 & 31 & 47441 & 30810 & 51508
\\ \cline{2-6}
 & 1212 & 59 & 46456 & 8043 & 59010
\\ \hline
\multirow{2}{*}{\expsmmm} & 1 & 6 & 42 & 25 & 42
\\ \cline{2-6}
 &  52 & 22 & 2077 & 530 & 2698
\\ \hline
\multirow{2}{*}{\expmttkrp} & 10 & 18 & 571 & 135 & 821
\\ \cline{2-6}
 &  239 & 35 & 8414 & 1130 & 10700
\\ \hline
\multirow{2}{*}{\expmmm} & 10 & 11 & 910 & 123 & 1242
\\ \cline{2-6}
 & 1708 & 61 & 33058 & 6479 & 43407
\\ \hline
\multirow{2}{*}{\expttm} & 11 & 12 & 1173 & 140 & 1480
\\ \cline{2-6}
 &  891 & 61 & 15891 & 3244 & 23981\\ \hline
    \end{tabular}
  \end{center}
  \end{small}
  \caption{Compilation metrics reported by Egg.}
  \label{tbl:egg:metrics}
  \vspace{-0.7cm}
\end{table}

\subsection{Optimization Overhead}
\label{exp:tradeoff}
In order to better demonstrate the practiciality of the optimization process,
we compare the run time and optimization time with the following coarse-grained rewrites: (1) storage-independent optimizations, and (2) optimizations that take storage into account. These two coincide with the optimization stages reported earlier.
As the tensor program, we consider the \expbatax{} kernel because (1) it has the longest
optimization overhead, and (2) it largely benefits from the two
stages of optimization.

\begin{figure}[t]
  \includegraphics[width=0.9\columnwidth]{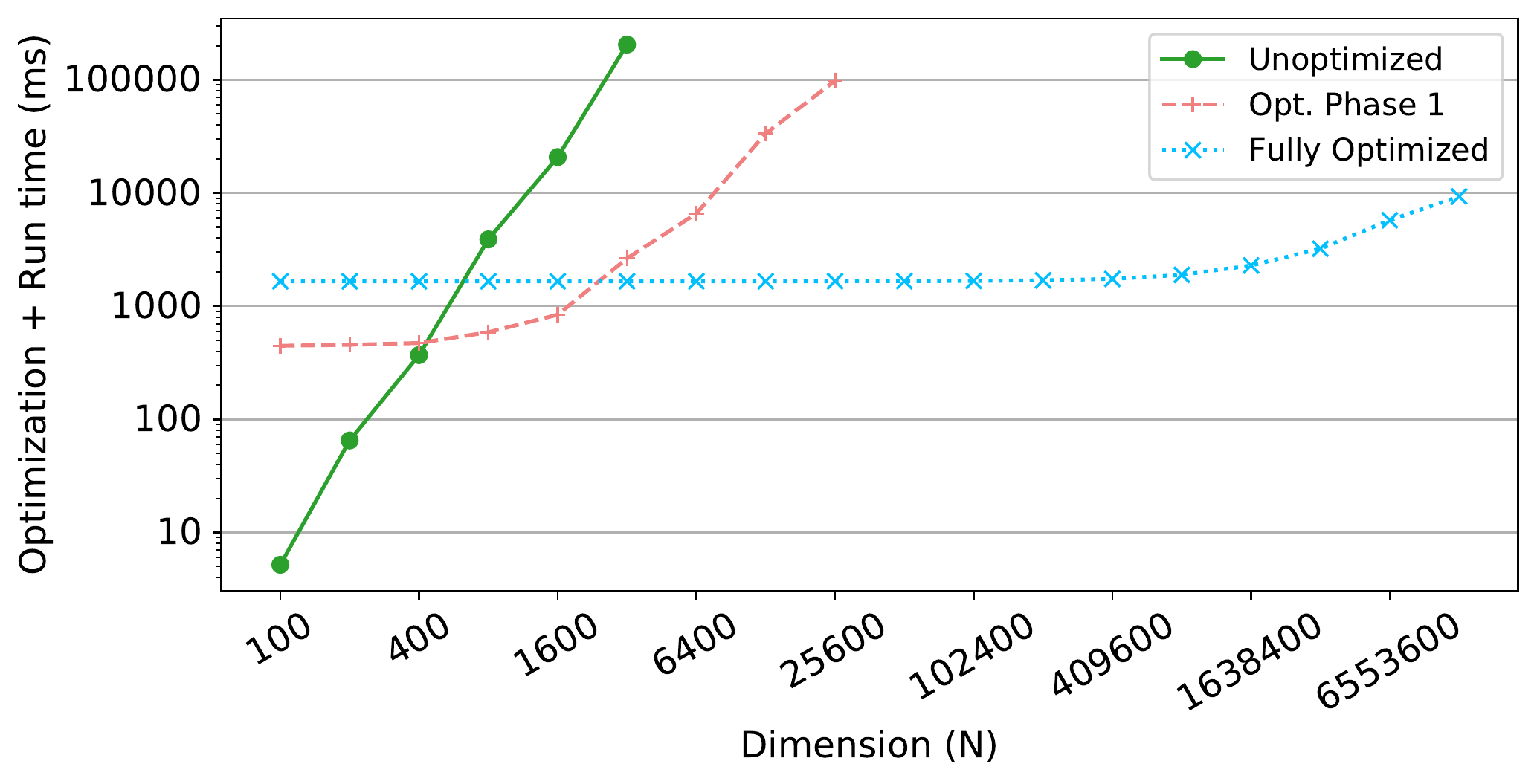}
  \vspace{-0.4cm}
  \caption{The total execution time of different versions of the \expbatax{} kernel. The dimension of matrix $A$ is $10^2\times N$.}
  \label{fig:exp:comp}
  \vspace{-0.2cm}
\end{figure}

Figure~\ref{fig:exp:comp} shows the total execution time of the \expbatax{} kernel,
including the optimization overhead,
by varying the dimension, for which we considered a time out of five minutes. We observe that although for smaller matrices
the unoptimized program is performing better,
for larger matrices the optimization overhead is amortized
by the improved run time.
The 1.7 seconds spent for the fully optimized kernel are well
justified.  They enable the system to scale to matrices that are three
orders of magnitude larger than those supported by the kernel with
only storage-independent optimizations.
Note that, while the optimization time is high, it
needs to be compared to compilers for tensor systems, which typically take much longer. For
instance, the BTO compiler can take several minutes to find the optimal execution
plan~\cite{bto}. In addition, Egg has been shown to outperform alternative approaches, such
as using SMT solvers~\cite{DBLP:journals/pacmpl/WillseyNWFTP21}.
In the next section, we
discuss the implications of these results.

\subsection{Discussion}
\label{sec:egg:discussion}

The use of the Equality Saturation System Egg was of great help for
us.  Egg supports the entire functionality needed for a rule
engine, it is an open source system, and it has been developed on
solid theoretical
foundations~\cite{DBLP:journals/pacmpl/WillseyNWFTP21}.  Nevertheless,
Egg is a research project, still under development, and has
limitations that affected our system \system; we discuss them here.

  {\em Performance} We used Egg version 0.6.0 and its optimizations adds significant overhead (c.f., Table~\ref{tbl:egg:metrics}). A very recent version was reported
recently~\cite{DBLP:journals/pacmpl/ZhangWWT22}, and it improves the matching significantly
by adopting a Worst Case Optimal Join~\cite{DBLP:journals/sigmod/NgoRR13}, since pattern
matching is, in essence, the same as computing a (usually cyclic) join query.  That version
is not yet available.

  {\em Cost computation} The biggest limitation for us is the way Egg handles the cost.  Egg
allows the user to define a cost model, and uses this cost model to extract the cheapest
expression from the root e-node.  However, it does not separate between the cardinality
estimate and the cost, and, worse, the cost can only be a number, while our cardinality,
defined in Sec.~\ref{sec:optimizer}, has a complex structure.  For that reason we had to use
hacks to approximate our cost using what is available in Egg.  We were able to always
extract the optimized plan for the given TSM, but were not able to compare in a meaningful way plans derived
from alternative storage formats, i.e. alternative TSMs. If this was possible, the programmer could  specify
several alternative storage mappings for one tensor. The system would then optimize the
program separately for each of them and return the cheapest plan.

  {\em Other minor limitations} The inability of Equality Saturation
Systems in general, and of Egg in particular, to handle expressions
with variables is well known.  In addition, a minor limitation is that
the current version of Egg does not have a DSL for the rules, instead
they need to be written in Rust.
We are in contact with the Egg authors and are optimistic that Egg
will continue to improve.

\balance

\section{Related Work}
\label{sec:related}

There is a vast literature on tensor and linear algebra systems in the compilers and HPC
communities. However, most of them focus on dense data
(e.g.,~\cite{DBLP:journals/pieee/PuschelMJPVSXFG05, spampinato2014basic,bto,halide,
chiw2012diderot, Steuwer:2015:GPP:2784731.2784754}). Similarly, the database community
studied Array DBMS~\cite{DBLP:journals/cse/StonebrakerBZB13, rasdaman} and SQL extensions
for matrices (e.g.,~\cite{DBLP:journals/cacm/LuoGGPJJ20, zhang2013sciql,
DBLP:conf/edbt/SchuleGK022}), which are also primarily designed for dense data.  Both
lines of work are not concerned with different tensor storage representations and thus
orthogonal to this work. Packages like \scipy{}~\cite{scipy} or the MATLAB Tensor
Toolbox~\cite{matlab:toolbox} support different sparse matrix/tensor representations, but
rely on composing hardcoded operations, which can become a severe bottleneck as shown in
Sec.~\ref{sec:experiments}. The closest related work from the tensor systems literature is
the Taco system~\cite{taco,taco-formats,taco-workspaces}, as highlighted in
Sec.~\ref{sec:intro}.

We drew many inspirations from the database literature.  At the top is the classic work on
GMAP~\cite{DBLP:journals/vldb/TsatalosSI96}, which pioneered the idea of using a declarative
language for representing physical data layout: for example, a secondary index can be
described as a view obtained by projecting the relation on the indexed attribute and the
primary key.  GMAP uses  Local As View (LAV), while our Tensor Storage Mappings are
defined as Global As View (GAV)~\cite{DBLP:journals/vldb/Halevy01}.  More recently the Hadad
system~\cite{hadad} has applied a similar high level principle for hybrid RA/LA analytics.
Hadad uses integrity constraints to express relationships between hybrid data sources, and
uses {\em chase} to optimize a query given those relationships.  The {\em chase} applies to
relational queries, over the Boolean domain, and does not extend to queries over semirings,
thus, it was not an option for our system.  The SPORES
system~\cite{DBLP:journals/pvldb/WangHSHL20} describes an optimizer for linear algebra, in
the context of SystemML~\cite{DBLP:journals/pvldb/BoehmRHSEP18}.  The key approach in SPORES
is to convert every query into a normal form, which is a sum of sum-of-products, i.e.
similar to Unions of Conjunctive Queries.  This is not possible in \lang, which we designed
specifically to cope with complex storage formats.  For example see the two quite different
expressions for matrix multiplication in Example~\ref{ex:matrix:multiplication}: there is no
unique normal form for that query.

Our work is also related to factorized learning, a line of work
  that uses database optimizations to improve the performance of
  machine learning tasks~\cite{DBLP:conf/sigmod/SchleichOK0N19,
    DBLP:conf/cgo/ShaikhhaSGO20, DBLP:journals/pvldb/ShaikhhaSO21,
    DBLP:journals/tods/KhamisNNOS20,
    DBLP:journals/pvldb/ChenKNP17,
    DBLP:conf/sigmod/KumarNP15}. Factorized learning, however,
  optimizes for normalized relational data; whereas we optimize for
  dense and sparse tensor representations. Normalized schemas are very
  different from COO/CSC/CSR/CSF representations.  The SystemML
  optimizer~\cite{DBLP:journals/pvldb/BoehmRHSEP18} has demonstrated
  the usefulness of loop fusion, an optimization that we capture with
  rule Rule F4 in Fig.~\ref{fig:opt_rules}.  Our optimizer is closest
  in spirit to SPORES~\cite{DBLP:journals/pvldb/WangHSHL20}, which
  optimizes linear algebra expressions by first converting them to
  relational algebra, optimizing these, then converting back to linear
  algebra.  The SPORES optimizer relies on the fact that the queries
  in that system have a unique normal form (since they are,
  essentially, UCQs).  Our optimization task is harder, because
  queries in \lang\ do not have a unique normal form, for example,
  consider the two matrix multiplication expressions in
  Example~\ref{ex:matrix:multiplication}, none of which can be
  considered to be the ``normal form'' of the other.

  At the time of writing, TensorFlow develops a graph optimization
  system called Grappler~\cite{grappler}; it is currently restricted
  to dense tensors, and is heuristic-based, while our system is
  cost-based.  A heuristic-based optimizer could, for example,
  prefer some physical plan when the tensors are dense, and another
  plan when they are sparse; in contrast, our cost-based optimizer can
  consider combinations of sparse and dense tensors and choose the
  most appropriate plan using a cost model.

\section{Conclusions}
\label{sec:conclusions}

We have described \system, which, to the best of our knowledge, is the first system to use a
cost-based optimizer to optimize tensor programs over flexible storage.  The key
contributions are the use of a common declarative language for both the Tensor Program and
the Tensor Storage Mappings, and a cost-based optimizer that can take advantage of rich
storage formats.  We have shown experimentally that the rule based optimizer can lead to
performance improvements over other systems. In future work, we plan to extend \system{} to
automatically choose between different storage formats. We also plan to integrate a
scheduler, which is inspired by the significant progress in automating the scheduler in
Halide~\cite{halide-autoschedule,halide-autoschedule-ml}.

  While our ultimate goal is to optimize entire ML pipelines,
  extending the current optimizer to large tensor programs will
  require significant engineering effort, and may also requires future
  research on how to propagate sparsity information of intermediate
  results.

\section*{Acknowledgement}
The authors would like to thank Remy Wang for his help with the Egg framework.
Shaikhha would like to thank Huawei for their support of the distributed data management and
processing laboratory at the University of Edinburgh.
Suciu was partially supported by NSF IIS 1907997 and NSF-BSF 2109922.
This project was partially supported by RelationalAI.


\bibliographystyle{ACM-Reference-Format}
\bibliography{main}

\end{document}